\pdfoutput=1
\documentclass[11pt,twoside,a4paper,cmspaper,final,collab]{cms-tdr}
\def\svnVersion{a9b510c-D}\def\svnDate{2021/07/13}\def\cmsCernNoTag{CERN-EP-2021-009}\def\cmsCernDate{\today}\def\cmsMessage{"Published in the European Physical Journal C as \href{http://dx.doi.org/10.1140/epjc/s10052-021-09348-6}{\doi{10.1140/epjc/s10052-021-09348-6}.}"}
\begin{document}\cmsNoteHeader{B2G-19-006}

\RCS$Revision$
\RCS$HeadURL$
\RCS$Id$

\newlength\cmsFigWidth
\newlength\cmsTabSkip\setlength{\cmsTabSkip}{1ex}
\ifthenelse{\boolean{cms@external}}{\setlength\cmsFigWidth{0.50\textwidth}}{\setlength\cmsFigWidth{0.65\textwidth}}
\ifthenelse{\boolean{cms@external}}{\providecommand{\cmsLeft}{upper\xspace}}{\providecommand{\cmsLeft}{left\xspace}}
\ifthenelse{\boolean{cms@external}}{\providecommand{\cmsRight}{lower\xspace}}{\providecommand{\cmsRight}{right\xspace}}
\ifthenelse{\boolean{cms@external}}{\providecommand{\cmsTable}[1]{#1}}{\providecommand{\cmsTable}[1]{\resizebox{\textwidth}{!}{#1}}}

\newcommand{\X}{{\HepParticle{X}{}{}}\xspace}
\newcommand{\mZpr}{\ensuremath{m_{\PZpr}}\xspace}
\newcommand{\VV}{\ensuremath{\PV\PV}\xspace}
\newcommand{\ZZ}{\ensuremath{\PZ\PZ}\xspace}
\newcommand{\WW}{\ensuremath{\PW\PW}\xspace}
\newcommand{\VH}{\ensuremath{\PV\PH}\xspace}
\newcommand{\gV}{\ensuremath{g_\text{V}}\xspace}
\newcommand{\cH}{\ensuremath{c_\text{H}}\xspace}
\newcommand{\cF}{\ensuremath{c_\text{F}}\xspace}
\newcommand{\ZprtoZH}{\ensuremath{\PZpr\to\PZ\PH}\xspace}
\newcommand{\Ztoll}{\ensuremath{\PZ\to\Pe\bar{\Pe},\Pgm\bar{\Pgm},\Pgt\bar{\Pgt}}\xspace}
\newcommand{\Ztonn}{\ensuremath{\PZ\to\nu\nu}\xspace}
\newcommand{\Wtoln}{\ensuremath{\PW\to\Pe\nu,\Pgm\nu,\Pgt\nu}\xspace}
\newcommand{\ZH}{\ensuremath{\PZ\PH}\xspace}
\newcommand{\mX}{\ensuremath{m_{\X}}\xspace}
\newcommand{\mtX}{\ensuremath{m_{\X}^{\text{T}}}\xspace}
\newcommand{\Vjets}{\ensuremath{\PV\text{+jets}}\xspace}
\newcommand{\Zjets}{\ensuremath{\PZ\text{+jets}}\xspace}
\newcommand{\mjj}{\ensuremath{m_{jj}}\xspace}
\newcommand{\jH}{\ensuremath{j_{\PH}}\xspace}
\newcommand{\mj}{\ensuremath{m_{\jH}}\xspace}
\newcommand{\ST}{single top quark\xspace}
\newcommand{\B}{\ensuremath{\mathcal{B}}}
\newcommand{\fb}{\ensuremath{\,\text{fb}}\xspace}

\cmsNoteHeader{B2G-19-006}
\title{Search for a heavy vector resonance decaying to a \PZ~boson and a Higgs boson in proton-proton collisions at \texorpdfstring{$\sqrt{s} = 13\TeV$}{sqrt(s) = 13 TeV}}
\titlerunning{Search for a heavy vector resonance decaying to Z + Higgs}
\date{\today}

\abstract{
A search is presented for a heavy vector resonance decaying into a \PZ boson and the standard model Higgs boson, where the \PZ boson is identified through its leptonic decays to electrons, muons, or neutrinos, and the Higgs boson is identified through its hadronic decays. The search is performed in a Lorentz-boosted regime and is based on data collected from 2016 to 2018 at the CERN LHC, corresponding to an integrated luminosity of 137\fbinv. Upper limits are derived on the production of a narrow heavy resonance \PZpr, and a mass below 3.5 and 3.7\TeV is excluded at 95\% confidence level in models where the heavy vector boson couples predominantly to fermions and to bosons, respectively. These are the most stringent limits placed on the Heavy Vector Triplet \PZpr model to date. If the heavy vector boson couples exclusively to standard model bosons, upper limits on the product of the cross section and branching fraction are set between 23 and 0.3\fb for a \PZpr mass between 0.8 and 4.6\TeV, respectively. This is the first limit set on a heavy vector boson coupling exclusively to standard model bosons in its production and decay.}

\hypersetup{
pdfauthor={CMS Collaboration},
pdftitle={Search for a heavy resonance decaying to a Z boson and a Higgs boson in proton-proton collisions at sqrt(s) = 13 TeV},
pdfsubject={CMS},
pdfkeywords={CMS, B2G, diboson, dilepton, HVT}
}

\maketitle

\section{Introduction}
The discovery of a Higgs boson (\PH)~\cite{bib:Aad20121,bib:Chatrchyan201230,Chatrchyan:2013lba} by the ATLAS and CMS Collaborations at the CERN LHC, with properties consistent with expectations from the standard model (SM) of particle physics, has emphasized the hierarchy problem of the SM. In the SM, the measured \PH mass of 125\GeV~\cite{Aad:2015zhl,Sirunyan_2020:diphoton}, given its fundamental scalar nature~\cite{Aad:2013xqa,Aad_2016}, requires extreme fine tuning of quantum corrections, suggesting that the SM may be incomplete. Many different exotic models, such as the little Higgs~\cite{Han:2003wu,Schmaltz,Perelstein2007247} and composite Higgs~\cite{Contino2011,Marzocca2012,Bellazzini:2014yua} models, predict the existence of new resonances decaying to a vector boson ($\PV = \PW, \PZ$) and a Higgs boson~\cite{1807.02826,1712.06518,Sirunyan:2019vgt,Aaboud:2018bun,Dorigo_2018}.

Heavy vector triplet (HVT) models~\cite{Pappadopulo2014} introduce new heavy vector bosons (\PWpr, \PZpr) that couple to the Higgs and SM gauge bosons with the parameters \cH and \gV, and to the fermions via the combination $(g^2/\gV) \cF$, where \cF is the fermion coupling and $g$ is the SM $\text{SU}(2)_\text{L}$ gauge coupling. The HVT couplings are expected to be of order unity in most models. Three benchmark models, denoted as models A, B, and C are considered in this paper.

In model A, the coupling strengths to fermions and gauge bosons are comparable and the heavy resonances decay predominantly to fermions, as is the case in some extensions of the SM gauge group~\cite{Barger:1980ix}. In model B, the fermionic couplings are suppressed, as in composite Higgs models. In model C, the fermionic couplings are set to zero, so the resonances are produced only through vector boson fusion (VBF) and decay exclusively to a pair of SM bosons. The parameters used for model A are $\gV = 1$, $\cH = -0.556$, and $\cF = -1.316$; for model B, $\gV = 3$, $\cH = -0.976$, and $\cF = 1.024$; and for model C, $\gV = 1$, $\cH = 1$, $\cF = 0$.

Previous searches for a heavy resonance decaying to a Higgs boson and a vector boson have been carried out at $\sqrt{s}= 13\TeV$ in the semileptonic final state~\cite{1808.01365,1807.02826,1712.06518} and in the fully hadronic final state~\cite{1707.06958,1707.01303,2007.05293} by the CMS and ATLAS Collaborations. The most stringent lower limit on the \PZpr mass at 95\% confidence level using the semileptonic (fully hadronic) final state is 2.65 (2.2)\TeV in HVT model A and 2.83 (2.65)\TeV in HVT model B~\cite{1712.06518,2007.05293}.

\begin{figure}[!htb]
  \centering
    \includegraphics[width=.35\textwidth]{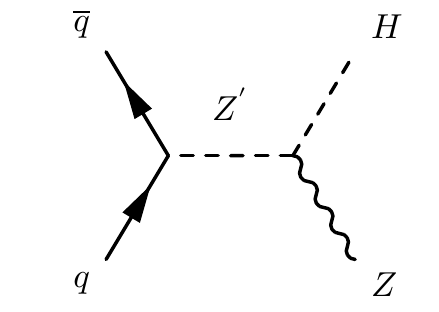}
    \includegraphics[width=.44\textwidth]{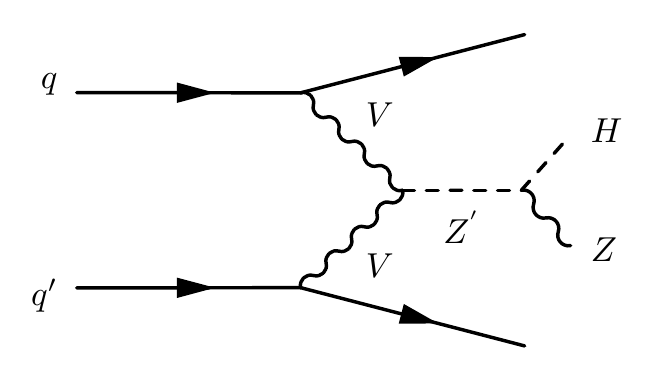}
  \caption{The leading order Feynman diagrams of the heavy resonance \PZpr production through \qqbar annihilation (\cmsLeft) and vector boson fusion (\cmsRight), decaying to a \PZ boson (\PZ) and a Higgs boson (\PH).}
  \label{fig:feynman}
\end{figure}

This paper describes a search for a heavy resonance (denoted as \PX for the reconstructed quantity and \PZpr for the particle predicted by the theory) decaying to a \PZ boson and a Higgs boson. The \PZ boson is identified via a pair of electrons or muons, or a large amount of missing transverse momentum (\ptvecmiss) measured in the detector due to the presence of at least two neutrinos. The Higgs boson is identified via its hadronic decays, either directly to a pair of heavy quarks, or via cascade decays dominated by \WW and \ZZ. We explore the regime where the Higgs boson has a large Lorentz boost and is reconstructed as a single, large-radius jet, referred to as \jH, with characteristic substructure and identified via its mass and possible presence of \PQb quark subjets. If a heavy resonance couples exclusively to the SM bosons, it can be produced dominantly through VBF. Dedicated categories are defined in order to enhance the sensitivity to this production mode, exploiting the presence of two jets with large transverse momenta (\pt) in the forward region of the detector, which are remnants of the initial-state quarks participating in the VBF interaction. The Feynman diagrams for the signal processes are depicted in Fig.\ref{fig:feynman}.

The search is performed by examining the distribution of the reconstructed mass (\mX) or transverse mass (\mtX) of the heavy resonance for a localized excess of events. The main background normalization is determined from data in sideband regions (SBs) of the \jH mass distribution, and extrapolated to the signal region (SR) through analytical functions derived from simulation.

\section{The CMS detector}

{\tolerance=800 The CMS detector features a silicon pixel and strip tracker, a lead tungstate crystal electromagnetic calorimeter (ECAL), and a brass and scintillator hadron calorimeter, each composed of a barrel and two endcap sections. These detectors reside within a superconducting solenoid, which provides a magnetic field of 3.8\unit{T}. Forward calorimeters extend the pseudorapidity $\eta$ coverage up to $\abs{\eta} < 5.2$. Muons are measured in gas-ionization detectors embedded in the steel flux-return yoke outside the solenoid. A detailed description of the CMS detector, together with a definition of the coordinate system and the kinematic variables, can be found in Ref.~\cite{Chatrchyan:2008zz}.\par}

Events of interest are selected using a two-tiered trigger system~\cite{Khachatryan:2016bia}. The first level, composed of custom hardware processors, uses information from the calorimeters and muon detectors to select events at a rate of around 100\unit{kHz} within a fixed time interval of about 4\mus. The second level, known as the high-level trigger (HLT), consists of a farm of processors running a version of the full event reconstruction software optimized for fast processing, and reduces the event rate to around 1\unit{kHz} before data storage. 

\section{Data and simulated samples}
The data samples used in this search were collected during the period 2016--2018, with the CMS detector at the LHC in proton-proton ($\Pp\Pp$) collisions at a center-of-mass energy of 13\TeV, resulting in a combined integrated luminosity of 137\fbinv.

{\tolerance=400 The signal samples are generated at leading order (LO) through \qqbar annihilation, taking the cross sections from HVT models A and B~\cite{Pappadopulo2014}, or through VBF with the cross section from HVT model C, using the \MGvATNLO 2.4.2~\cite{bib:MADGRAPH} generator and the MLM matching scheme~\cite{bib:MLM}. Different hypotheses for the heavy resonance mass in the range of 800 to 5000\GeV are considered, with the natural width of the resonance being negligible compared to the 4\% detector resolution (the narrow-width approximation). The heavy resonance is forced to decay to a \PZ boson and a Higgs boson, with the former decaying into a pair of charged leptons ($\ell = \Pe$ or $\PGm$) or neutrinos, including cascade decays involving tau leptons. There is no restriction on the decay channels for the Higgs boson and its decay particles, which decay according to the SM branching fractions. \par}

{\tolerance=800 The SM background for this search is dominated by \Vjets production, with the \PV boson decaying as \Ztonn, \Ztoll, or \Wtoln. The \Vjets background sample is produced with the \MGvATNLO generator at LO. The sample is further normalized to account for next-to-LO (NLO) in electroweak (EW) and next-to-NLO (NNLO) in quantum chromodynamics (QCD) corrections to the cross section from Ref.~\cite{Lindert:DYCorrections}. The top quark pair (\ttbar) and single top quark $t$-channel and $\cPqt\PW$ production are generated at NLO in QCD with the \POWHEG 2.0 generator~\cite{Nason:2004rx,Frixione:2007vw,Alioli:2010xd,Frederix_2012,Re_2011,Campbell_2015}. The \ttbar samples are normalized to the cross section computed with \textsc{Top++} 2.0~\cite{Czakon:2011xx} at NNLO in QCD with next-to-next-to-leading logarithmic soft gluon resummation accuracy. The single top quark $s$-channel, \VV, and \VH samples are simulated at NLO in QCD with the \MGvATNLO generator. \par}

The NNPDF 3.0~\cite{Ball_2015} set of parton distribution functions (PDF) is used to simulate the hard process in all simulated samples for the 2016 data and the NNPDF 3.1~\cite{Ball_2017} set is used for 2017 and 2018. Parton showering and hadronization processes are performed with  \PYTHIA 8.226~\cite{Sjostrand_2015} with the CUETP8M1~\cite{Skands:2014pea,Khachatryan:2015pea} underlying event tune for 2016, and \PYTHIA 8.230 with the CP5~\cite{pythia8_tuneCP5} event tune for 2017 and 2018. The CUETP8M2 underlying event tune~\cite{CMS-PAS-TOP-16-021} is used to simulate \ttbar production for 2016 samples. The CMS detector response simulation is performed with \GEANTfour~\cite{bib:GEANT4}. Simulated samples are reconstructed with the same software as used for collision data. The data samples contain additional $\Pp\Pp$ interactions in the same or nearby bunch crossings (pileup). The simulated pileup description is reweighted to match the distribution of the pileup multiplicity measured in data.

\section{Event reconstruction}
Events in the CMS detector are reconstructed using the particle-flow (PF) algorithm~\cite{bib:PF1}, which combines information from all subdetectors in order to reconstruct stable particles (muons, electrons, photons, neutral and charged hadrons). Jets are reconstructed from PF candidates clustered with the anti-\kt algorithm~\cite{Cacciari:2008gp}, with a distance parameter of 0.4 (AK4 jets) or 0.8 (AK8 jets), using the {\FASTJET} 3.0 package~\cite{Cacciari:2011ma,Cacciari:2008gn}. Several vertices are reconstructed per bunch crossing. The candidate vertex with the largest value of summed physics-object $\pt^2$ is taken to be the primary $\Pp\Pp$ interaction vertex. Here the physics objects are the AK4 jets, clustered using the jet finding algorithms with the tracks assigned to candidate vertices as inputs, and the associated \ptvecmiss taken as the negative vector \pt sum of those jets. Two different methods to remove contributions from pileup are used: for the AK4 jets, pileup is accounted for via the charged-hadron subtraction algorithm~\cite{Sirunyan_2020} in conjunction with the jet area method~\cite{Cacciari_2008}, while for the AK8 jets the pileup-per-particle identification algorithm~\cite{Bertolini2014} is employed. The jet energy resolution, after the application of corrections to the jet energy, is 4\% at 1\TeV~\cite{Khachatryan:2016kdb}. For the AK4 jets, $\pt > 30\GeV$ and $\abs{\eta} < 2.4$ are required, and jets within a cone of $\Delta R(j,\ell)=\sqrt{\smash[b]{\Delta\eta(j,\ell)^2+\Delta\phi(j,\ell)^2}}>0.4$ around isolated leptons are removed, where $\phi$ is the azimuthal angle. The AK8 jets must satisfy $\pt > 200$\GeV and $\abs{\eta} < 2.4$. The vector~\ptvecmiss is computed as the negative vector \pt sum of all the PF candidates in an event. The \ptvecmiss is corrected for adjustments to the energy scale of the reconstructed AK4 jets in the event, and its magnitude is denoted as \ptmiss~\cite{Sirunyan_2019}. The observable \mht is defined as the magnitude of the vector \pt sum of all AK4 jets with $\pt > 30\GeV$ and $\abs{\eta} < 3.0$.

For each AK8 jet a groomed jet mass ($m_j$) is calculated, after applying a modified mass-drop algorithm~\cite{Dasgupta:2013ihk,Butterworth:2008iy}. The mass-drop algorithm used here is known as the soft-drop algorithm~\cite{Larkoski:2014wba}, with parameters $\beta=0$, $z_\text{cut}=0.1$, and $R_0 = 0.8$. Subjets are obtained by reverting the last step of the jet clustering and selecting the two with the highest \pt. The groomed jet mass is calibrated in a \ttbar sample enriched in hadronically decaying \PW bosons~\cite{Khachatryan:2014vla}.

The identification of jets that originate from \PQb quarks is performed with the DeepCSV algorithm~\cite{DeepCSV}, which is based on a deep neural network with information on tracks and secondary vertices associated with the jet as inputs. The DeepCSV algorithm is applied to AK4 jets and the two highest \pt AK8 subjets. A jet is considered as \PQb tagged if the output discriminator value is larger than a defined threshold, corresponding to a 75\% \PQb tagging efficiency with a probability for mistagging jets originating from the hadronization of gluons or \PQu/\PQd/\PQs quarks of about 3\%. The simulated samples are reweighted to account for small differences in the \PQb tagging efficiency from values obtained in data.

Electrons are reconstructed from ECAL energy deposits in the range $\abs{\eta}<2.5$ that are matched to tracks reconstructed in the silicon tracker. The electrons are identified taking into account the distribution of energy deposited along the electron trajectory, the direction and momentum of the track, and its compatibility with the primary vertex~\cite{Khachatryan:2015hwa}. Electrons are required to pass an isolation requirement. The isolation is defined as the \pt sum of all particles within a cone of $\Delta R = 0.3$ around the electron track, after the contributions from the electron itself, other nearby electrons, and pileup are removed. The electron reconstruction efficiency is larger than 88\%.

Muons are reconstructed within the acceptance of $\abs{\eta} < 2.4$ by matching tracks in the silicon tracker and charge deposits (hits) in the muon spectrometer. Muon candidates are identified via selection criteria based on the compatibility of tracks reconstructed from only silicon tracker information with tracks reconstructed from a combination of the hits in both the tracker and muon detector. Additional requirements are based on the compatibility of the trajectory with the primary vertex, and on the number of hits observed in the tracker and muon systems. Muons are required to be isolated by imposing a limit on the \pt sum of all the reconstructed tracks within a cone $\Delta R = 0.4$ around the muon direction, excluding the tracks attributed to muons, divided by the muon \pt. The efficiency to reconstruct and identify muons is larger than 96\%~\cite{Sirunyan:2018}.

Hadronically decaying $\tau$ leptons (\tauh) are reconstructed by combining one or three charged particles with up to two neutral pion candidates. The selection criteria for the \tauh candidates, which are used to veto various backgrounds, are $\pt > 18\GeV$, $\abs{\eta} < 2.3$, and $\Delta R > 0.4$, where $\Delta R$ is a candidate's separation from isolated electrons and muons in the event~\cite{tau_reco}.

\section{Event selection}

Events are divided into categories depending on the number and flavor of the reconstructed leptons, the number of \PQb-tagged subjets of the Higgs candidate jet (\jH), and the presence of forward jets consistent with originating from VBF processes. In total, 12 categories are defined and listed in Table~\ref{tab:cat}.

\begin{table}[!htb]
  \topcaption{List of the 12 event categories used in the analysis.}\label{tab:cat}
  \centering
  \setlength{\tabcolsep}{16pt}
    \begin{tabular}{ll}
    		$0\ell$, 2\PQb tag, non-VBF & $0\ell$, 2\PQb tag, VBF\\
    		$2\Pe$, 2\PQb tag, non-VBF & $2\Pe$, 2\PQb tag, VBF\\
    		$2\PGm$, 2\PQb tag, non-VBF & $2\PGm$, 2\PQb tag, VBF\\
    		$0\ell$, $\leq$1\PQb tag, non-VBF & $0\ell$, $\leq$1\PQb tag, VBF\\
    		$2\Pe$, $\leq$1\PQb tag, non-VBF & $2\Pe$, $\leq$1\PQb tag, VBF\\
    		$2\PGm$, $\leq$1\PQb tag, non-VBF & $2\PGm$, $\leq$1\PQb tag, VBF\\
    \end{tabular}
\end{table}

The highest \pt AK8 jet in the event is assigned to \jH, and is required to have a transverse momentum $\pt^{\PH}> 200\GeV$ and $\abs{\eta} < 2.4$. This is the correct jet choice in 96\% of the simulated signal events. The minimal separation between \jH and isolated leptons from the \PZ boson decay is required to satisfy $\Delta R(\jH,\ell) > 0.8$. The mass of the \jH jet is required to be compatible with the \PH mass ($105<\mj<135\GeV$). It can have 0, 1, or 2 subjets that pass the \PQb tagging selection. If both subjets are \PQb tagged, the event belongs to the 2\PQb tag category, otherwise it is assigned to the $\leq$1\PQb tag category.

{\tolerance= 800
The $0\ell$ categories require $\ptmiss > 250\GeV$, originating from the Lorentz-boosted \PZ boson decaying to two neutrinos, which leave the detector unobserved. Data are collected using trigger selections that require $\ptmiss > 110\GeV$, calculated with or without considering muons, or $\mht > 110\GeV$. The minimal azimuthal angular separation between all AK4 jets and the \ptvecmiss vector has to satisfy $\Delta\phi(j,\ptvecmiss) > 0.5$ in order to suppress multijet production. The azimuthal angular separation between \jH and \ptvecmiss must satisfy $\Delta\phi(\jH,\ptvecmiss) > 2$. Events arising from detector noise are removed by requiring that the fractional contribution of charged hadron candidates to the \PH momentum be larger than 0.1, and the ratio $\ptmiss/\pt^\PH$ be larger than 0.6. Events with isolated leptons with $\pt > 10\GeV$ or hadronically decaying $\tau$ leptons with $\pt > 18$\GeV are removed in order to reduce the contribution from other SM processes. The \ttbar contribution is reduced by removing events with an additional \PQb-tagged AK4 jet not overlapping with \jH such that $\Delta R(j,\jH)>1.2$ is satisfied. Since the resonance mass cannot be reconstructed because of the presence of undetected decay products, the \jH momentum and the \ptvecmiss are used to compute the transverse mass $\mtX = \sqrt{\smash[b]{2\ptmiss \pt^\PH(1-\cos\Delta\phi(\ptvecmiss,\ptvec^\PH))}}$. In the VBF category, the condition $\abs{\eta_{j_\PH}}<1.1$ is applied on the $\jH$ to reduce the contribution of events where the measured $\mtX$ is significantly below $\mZpr$.
\par}

For the 2\Pe categories, data are collected using an electron trigger that requires either an isolated electron with $\pt > 35\GeV$ or a nonisolated electron with $\pt > 115\GeV$. In the 2\PGm categories, a muon trigger that requires a nonisolated muon with $\pt > 50\GeV$ is used to collect data. For both the 2\Pe and 2\PGm categories, the two selected leptons must have opposite charge, $\pt > 55$ and 20\GeV, respectively, and should be isolated from other activity in the event, except for each other. The \PZ boson candidates are required to have a dilepton invariant mass in the range 70--110\GeV, and $\pt > 200\GeV$. The \PZ boson mass window is large compared with the dilepton mass resolution, which is 3 (4)\% for an electron (muon) pair. A more stringent selection would decrease both the signal and the \Zjets background selection efficiency by the same amount, thus reducing the signal sensitivity. The separation between the \PZ boson candidate and \jH is required to be $\Delta R(\jH,Z) > 2$ for all categories, and $\abs{\Delta\eta(\jH,Z)} < 1.7$ additionally for the non-VBF categories, to further reduce the \Zjets background.

Candidate VBF events are selected in both the $0\ell$ and $2\ell$ categories by requiring two additional AK4 jets (j) with $\abs{\eta_j}<5$ that satisfy $\Delta R(j,\jH)>1.2$ in order to avoid overlap with the \jH, have $\eta_j$ values of opposite sign, a dijet mass $\mjj > 500\GeV$, and that satisfy a separation $\Delta\eta_{jj} > 4$. The two AK4 jets with the highest dijet mass are selected.

\begin{figure*}[!htb]
  \centering
    \includegraphics[width=0.45\textwidth]{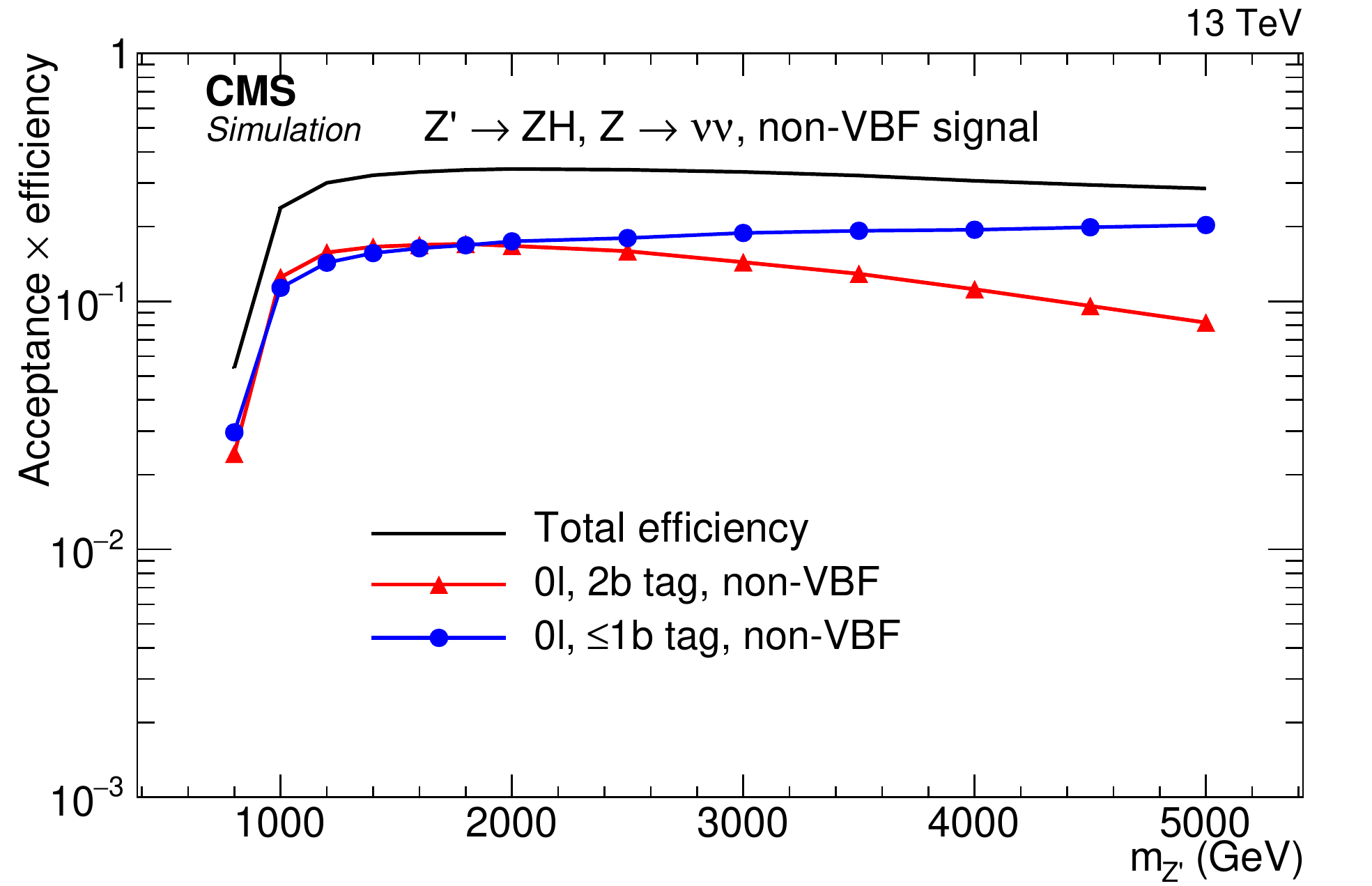}
    \includegraphics[width=0.45\textwidth]{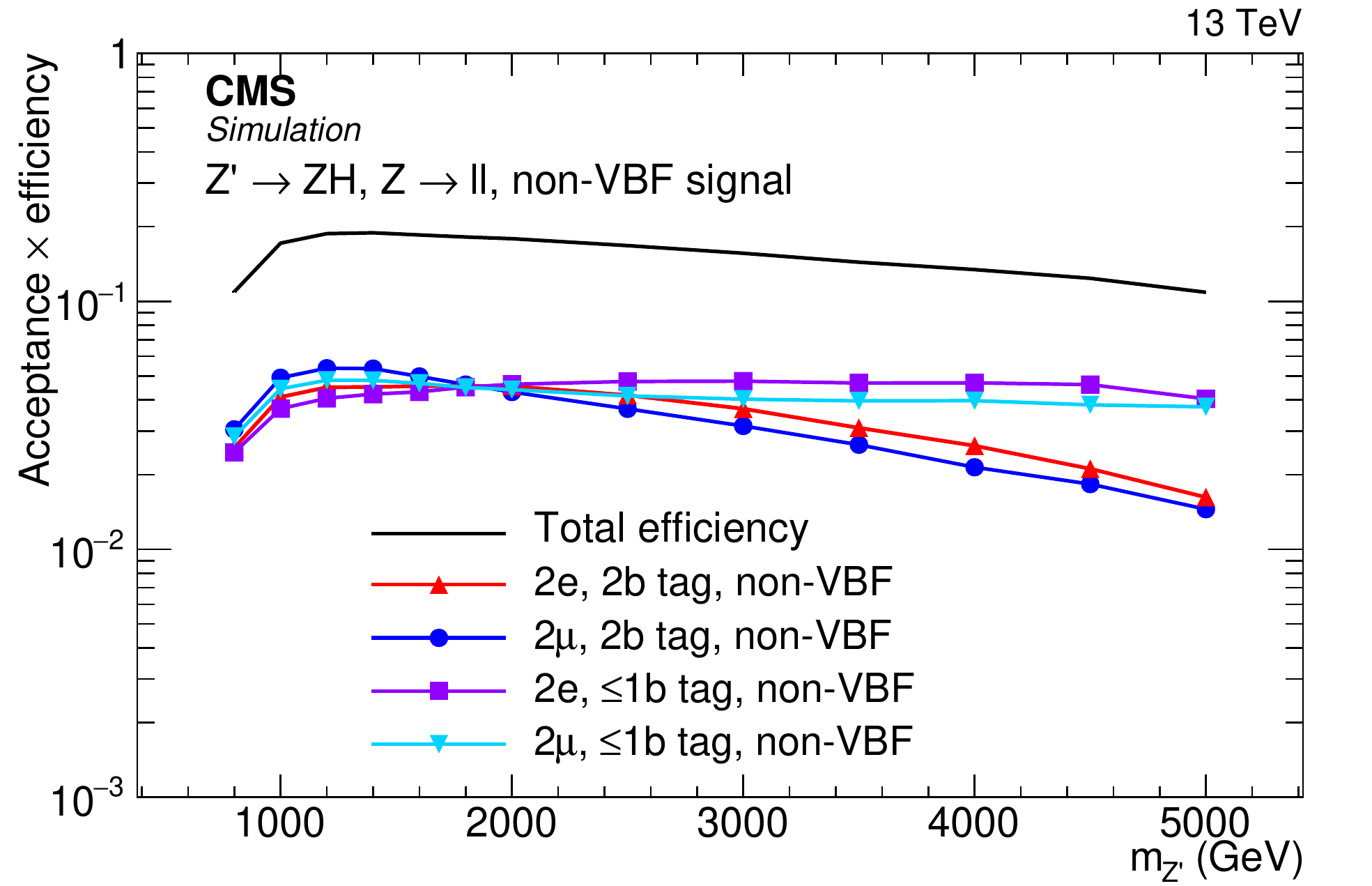}
    \includegraphics[width=0.45\textwidth]{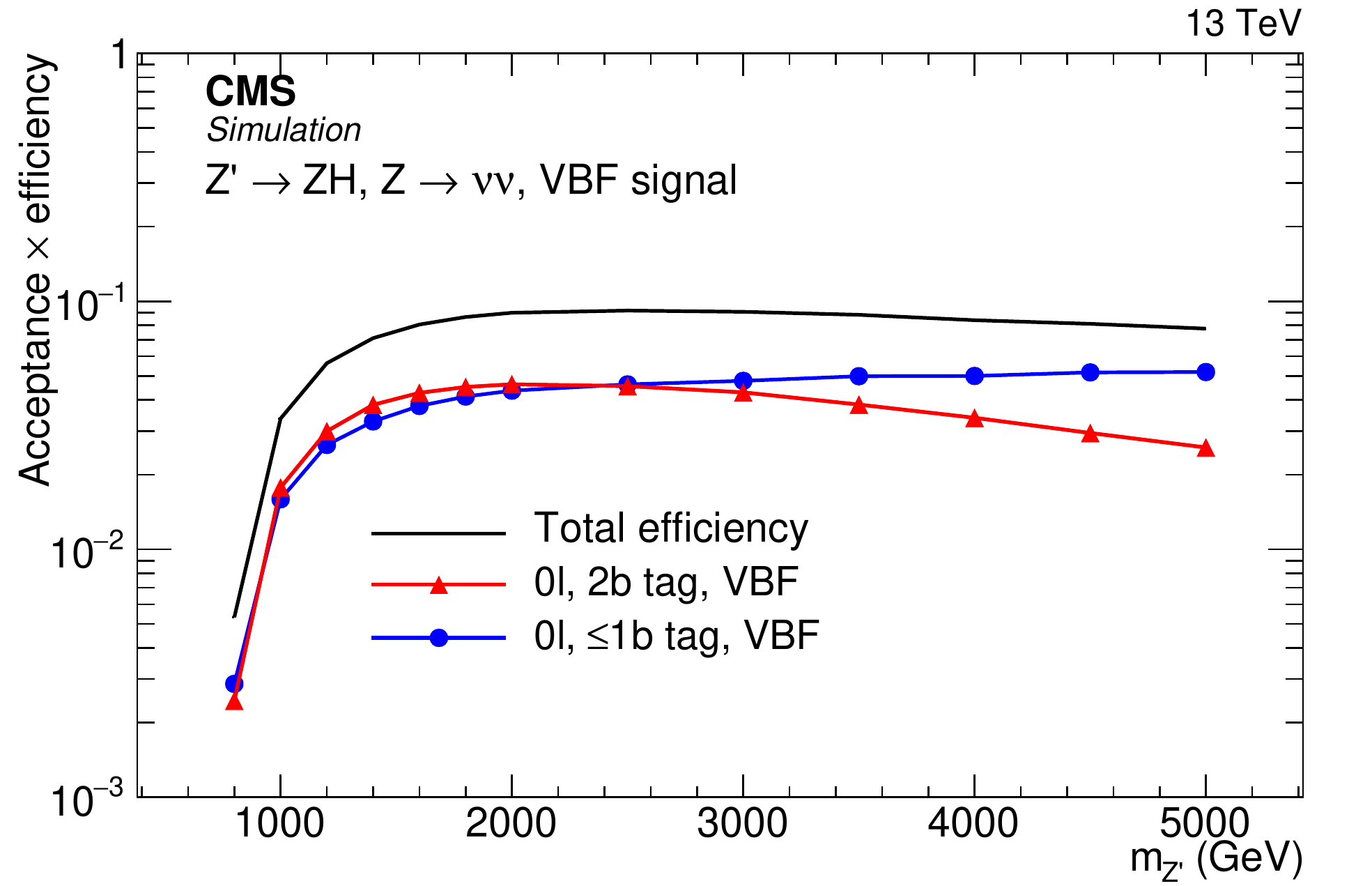}
    \includegraphics[width=0.45\textwidth]{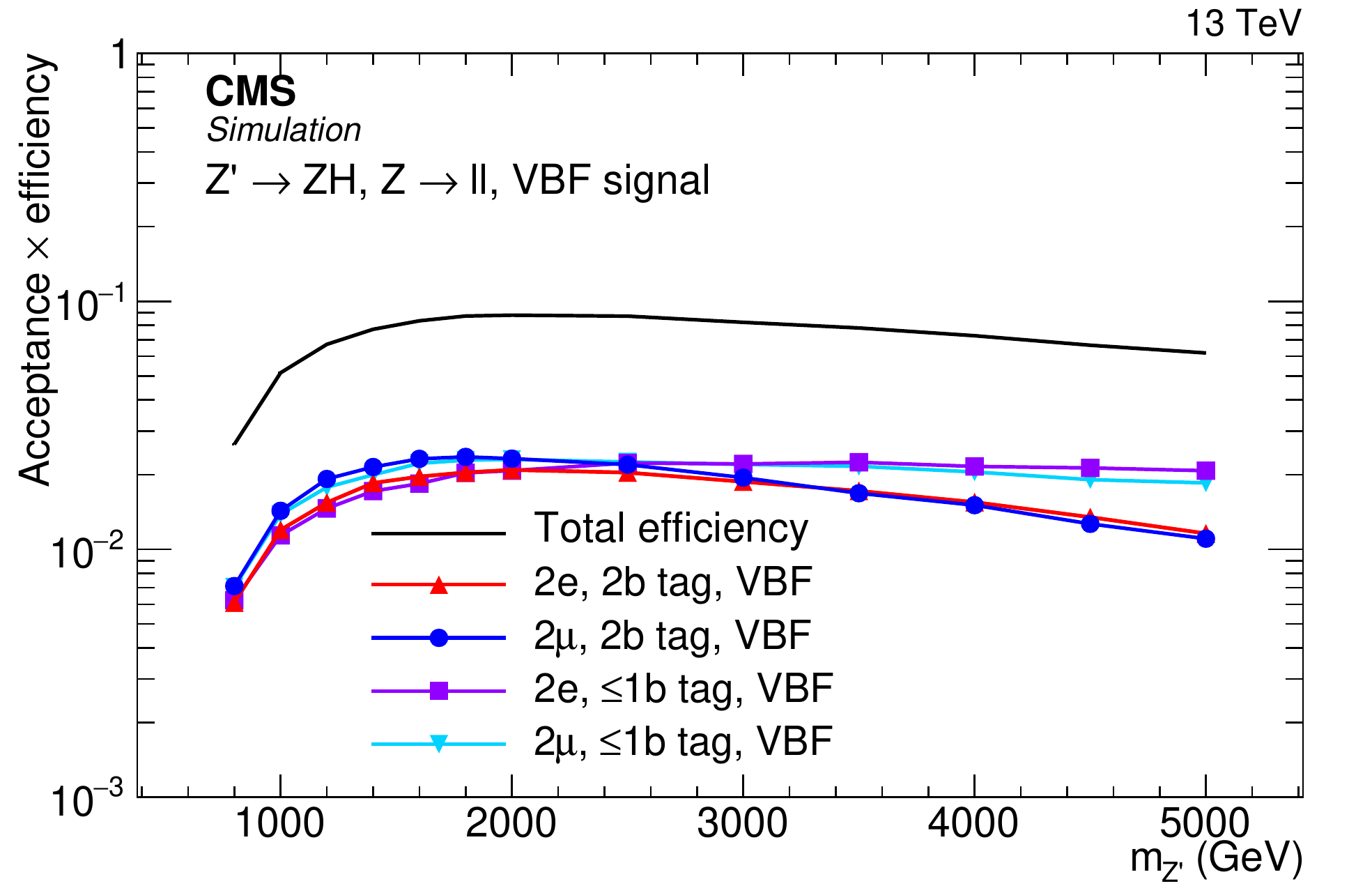}
    \caption{The product of signal acceptance and efficiency in the $0\ell$ (left column) and $2\ell$ (right column) categories for the signal produced via \qqbar annihilation (upper row) and vector boson fusion (lower row).}			\label{fig:signal_eff}
\end{figure*}

A further requirement is to have either \mX or \mtX larger than 1200\GeV for the $\leq$1\PQb tag, non-VBF categories, and larger than 750\GeV for the other categories to ensure the smoothness of the background model. The product of the signal geometrical acceptance and the selection efficiency, reported in Fig.~\ref{fig:signal_eff}, is calculated for the $0\ell$ category with the denominator being the \PZ decay to neutrinos, and for the $2\ell$ categories with the denominator being the \PZ decay to electrons, muons and tau leptons.

\section{Background estimation and signal modeling}
The most important SM background is vector boson production in association with \PQb-tagged jets (\Vjets). The \Vjets background is estimated using control samples in data to reduce the dependence on simulation. Minor SM backgrounds are \ttbar and single top quark processes, SM diboson production (\VV), and SM \PH production in association with a vector boson (\VH), all of which are estimated based on simulation. The SM \ZH production is considered as a background in this analysis. However, this process can be distinguished from the signal because of the non-resonant distribution in the \ZH invariant mass and by the softer \pt spectra of the \PH and \PZ bosons. The jet mass distribution is split into a signal-enriched region (SR) with $105<\mj<135\GeV$, and low-mass and high-mass sidebands (SB) with $30<\mj<65\GeV$ (LSB) and $135<\mj<250\GeV$ (HSB), respectively. The jet mass range $65<\mj<105\GeV$, a region enriched with boosted vector bosons (VR), is excluded and kept blinded in order to avoid potential contamination from a \VV resonant signal, which is the subject of dedicated searches~\cite{XtoVVnnqq,XtoVVllqq,Sirunyan:2019vgt}. The background estimation consists of two separate steps to determine, first, the number of events and, second, the distribution of the main background in the SR.

\subsection{Background normalization}
\label{subsec:norm}
The three groups of backgrounds (\Vjets, \ttbar and single top quark, and \VV and \VH) are considered separately, since each group has different physical properties leading to a different shape of the jet mass distribution. An appropriate analytical function is chosen to describe the background in each case. The \Vjets background's Higgs candidate jet mass has a smoothly falling shape with no peaks, therefore Chebyshev polynomials of order 1--4 are chosen to model the distribution observed in data. The \VV and \VH backgrounds have two peaks in the jet mass distribution, corresponding to the \PW and \PZ bosons, and the \VH background an additional peak due to the Higgs boson. The \ttbar and single top quark backgrounds are considered together, because they both have two peaks corresponding to $\PW \to \qqbar'$ decays and all-hadronic top quark decays $\cPqt \to \PW\PQb \to \qqbar'\PQb$.

The normalization of the simulated top quark background is corrected with a scale factor (SF) determined in high-purity top quark control regions. In the $0\ell$ category, the control region is defined by the veto on the additional \PQb-tagged AK4 jet being inverted. In the $2\ell$ categories, control region data are collected using the same trigger as for the 2\Pe signal region, with a requirement that lepton flavors and charges are different, resulting in a $1\Pe1\PGm$ region, where the leptons must have a combined invariant mass $m_{\Pe\PGm} > 110$\GeV and a vector sum $\pt^{\Pe\PGm} > 120$\GeV. Multiplicative SFs are calculated from the ratio of the event yield between data and simulation and are applied to the simulated samples in the SR. The uncertainties in the top quark SFs originate from the limited event count in the top quark control region and the extrapolation from the top quark control region to the SR. The systematic uncertainty in the $0\ell$ category is derived by varying the \PQb tagging SF. For the $2\ell$ categories the uncertainties in the electron and muon identification are taken into account. The electron and muon trigger uncertainties only affect the 2\PGm and not the 2\Pe category because the electron trigger is used to provide the control region while the muon trigger is used to select the signal region. A normalization uncertainty is applied to the VBF categories to account for the limited event counts in these control regions. The normalization uncertainty is taken as the deviation of the top quark SF from unity as shown in Table~\ref{tab:TopCR}.

\begin{table*}[!htb]
  \topcaption{Scale factors derived for the normalization of the \ttbar and single top quark backgrounds for different event categories. Uncertainties due to the limited size of the event samples (stat.) and systematic effects (syst.) are reported as well. The scale factors of the 2\Pe and 2\PGm categories are derived using the $1\Pe1\PGm$ top quark control region as described in the text.}
  \centering
    \begin{tabular}{cccccc}
    	  \hline
      \multicolumn{2}{c}{Non-VBF category} & \ttbar, \ST \,\,SF $\pm$ stat. $\pm$ syst.\\
      \hline
      \multirow{3}{*}{2\PQb tag} & $0\ell$ & 1.012 $\pm$ 0.116 $\pm$ 0.008\\
      & $2\Pe$ & 1.098 $\pm$ 0.084 $\pm$ 0.067\\
      & $2\PGm$ & 1.098 $\pm$ 0.084 $\pm$ 0.075\\[\cmsTabSkip]
      \multirow{3}{*}{$\leq$1\PQb tag} & $0\ell$ & 1.028 $\pm$ 0.048 $\pm$ 0.009\\
      & $2\Pe$ & 1.003 $\pm$ 0.021 $\pm$ 0.089\\
      & $2\PGm$ & 1.003 $\pm$ 0.021 $\pm$ 0.095\\[\cmsTabSkip]
      \hline
      \multicolumn{2}{c}{VBF category} & \ttbar, \ST \,\,SF $\pm$ stat. $\pm$ syst. $\pm$ VBF norm.\\
      \hline
      \multirow{3}{*}{2\PQb tag} & $0\ell$ & 0.676 $\pm$ 0.221 $\pm$ 0.007 $\pm$ 0.330\\
      & $2\Pe$ & 0.676 $\pm$ 0.154 $\pm$ 0.096 $\pm$ 0.330\\
      & $2\PGm$ & 0.676 $\pm$ 0.154 $\pm$ 0.103 $\pm$ 0.330\\[\cmsTabSkip]
      \multirow{3}{*}{$\leq$1\PQb tag} & $0\ell$ & 0.822 $\pm$ 0.144 $\pm$ 0.022 $\pm$ 0.180\\
      & $2\Pe$ & 0.882 $\pm$ 0.044 $\pm$ 0.099 $\pm$ 0.120\\
      & $2\PGm$ & 0.882 $\pm$ 0.044 $\pm$ 0.107 $\pm$ 0.120\\
      \hline
    \end{tabular}
    \label{tab:TopCR}
\end{table*}

\begin{figure*}[!htbp]
  \centering
    \includegraphics[width=.42\textwidth]{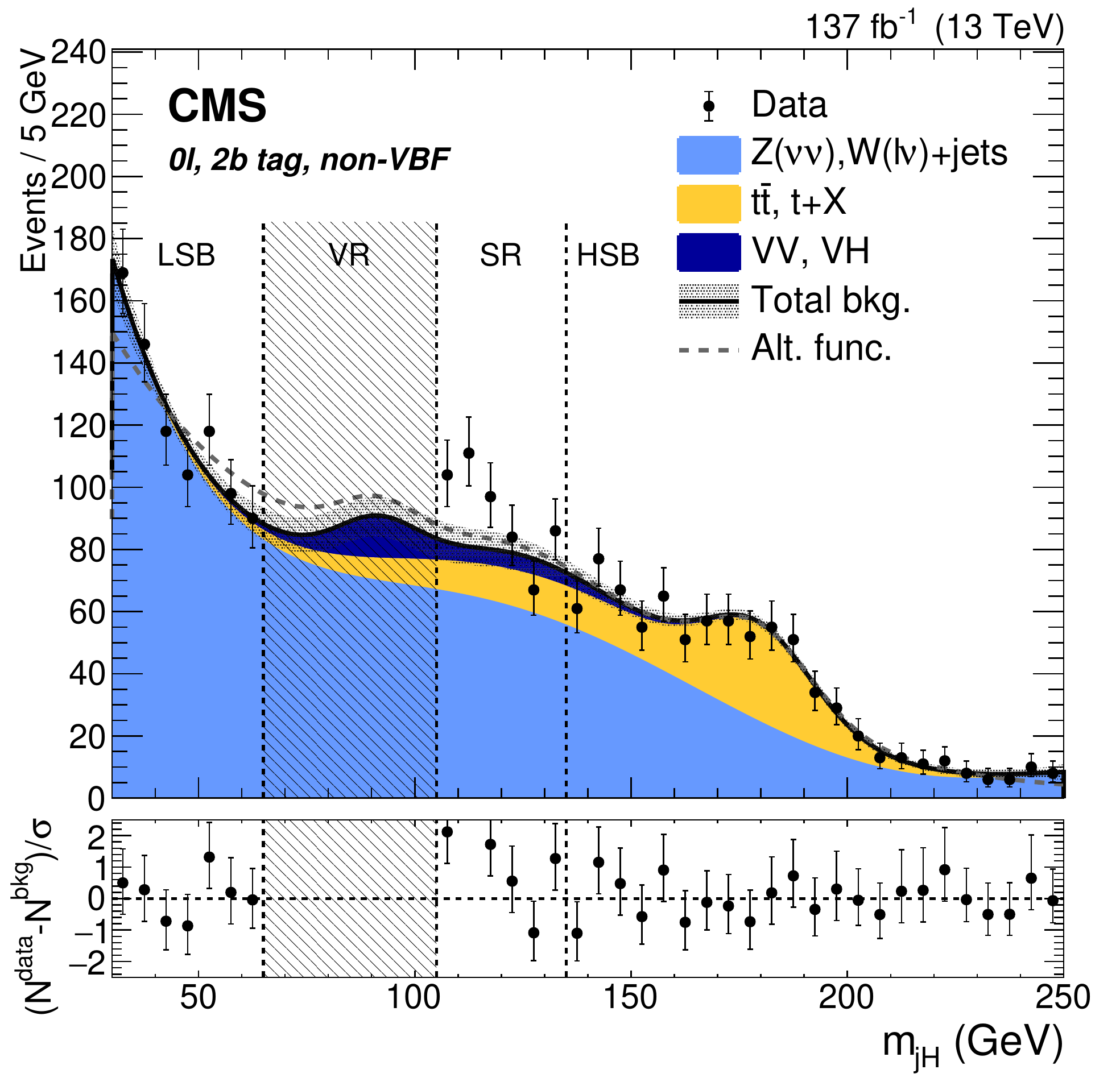}
    \includegraphics[width=.42\textwidth]{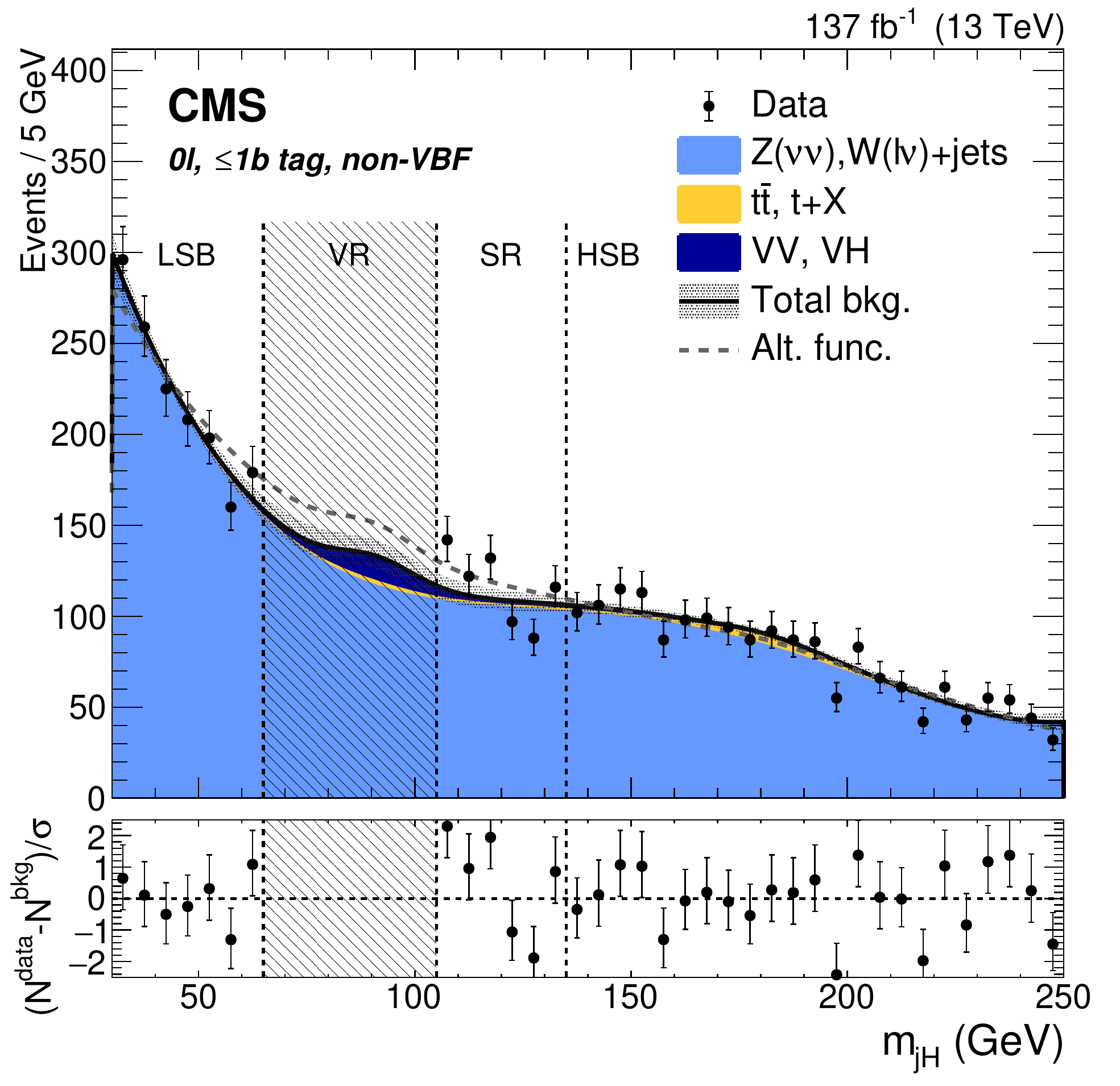}
    \includegraphics[width=.42\textwidth]{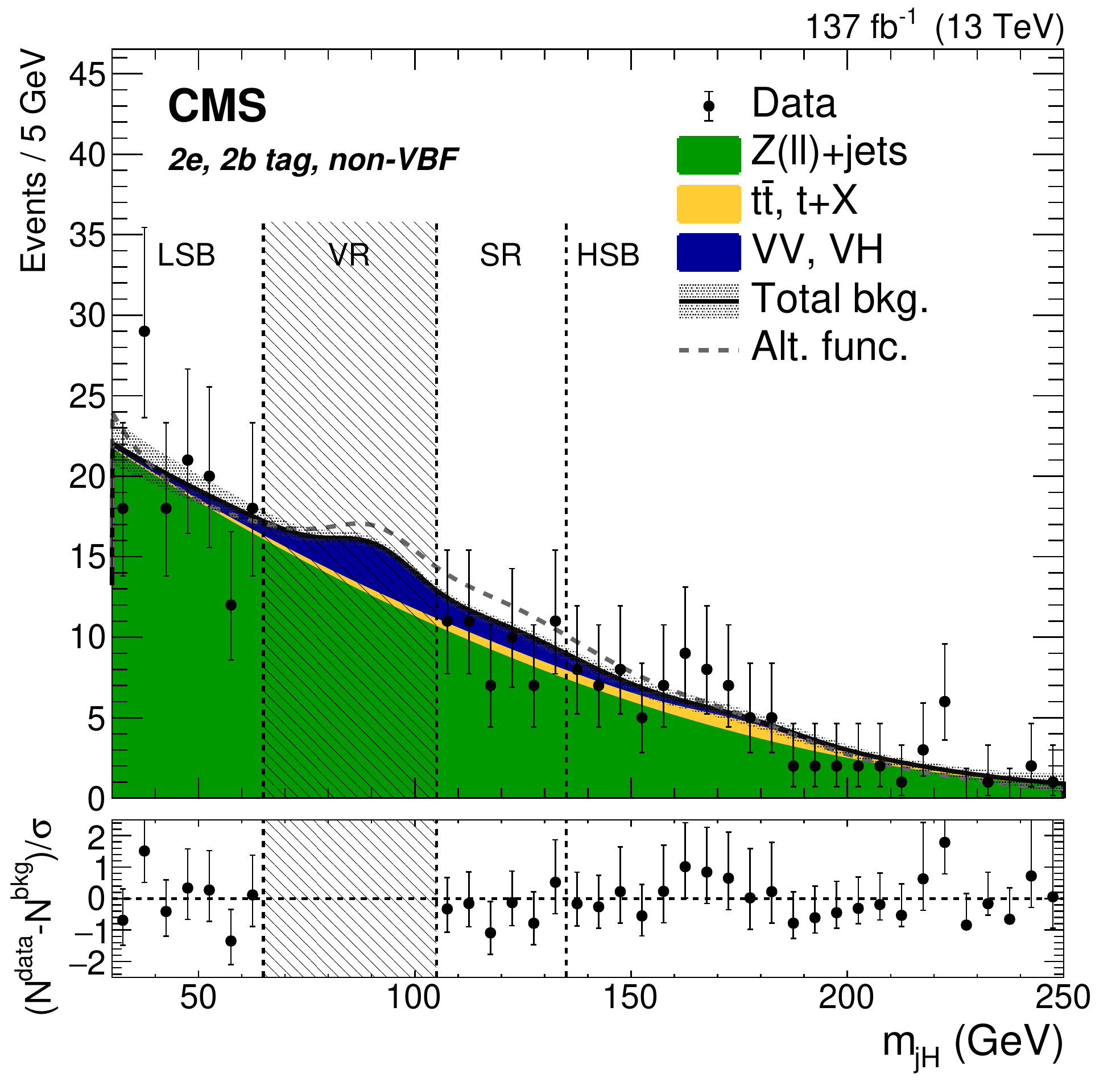}
    \includegraphics[width=.42\textwidth]{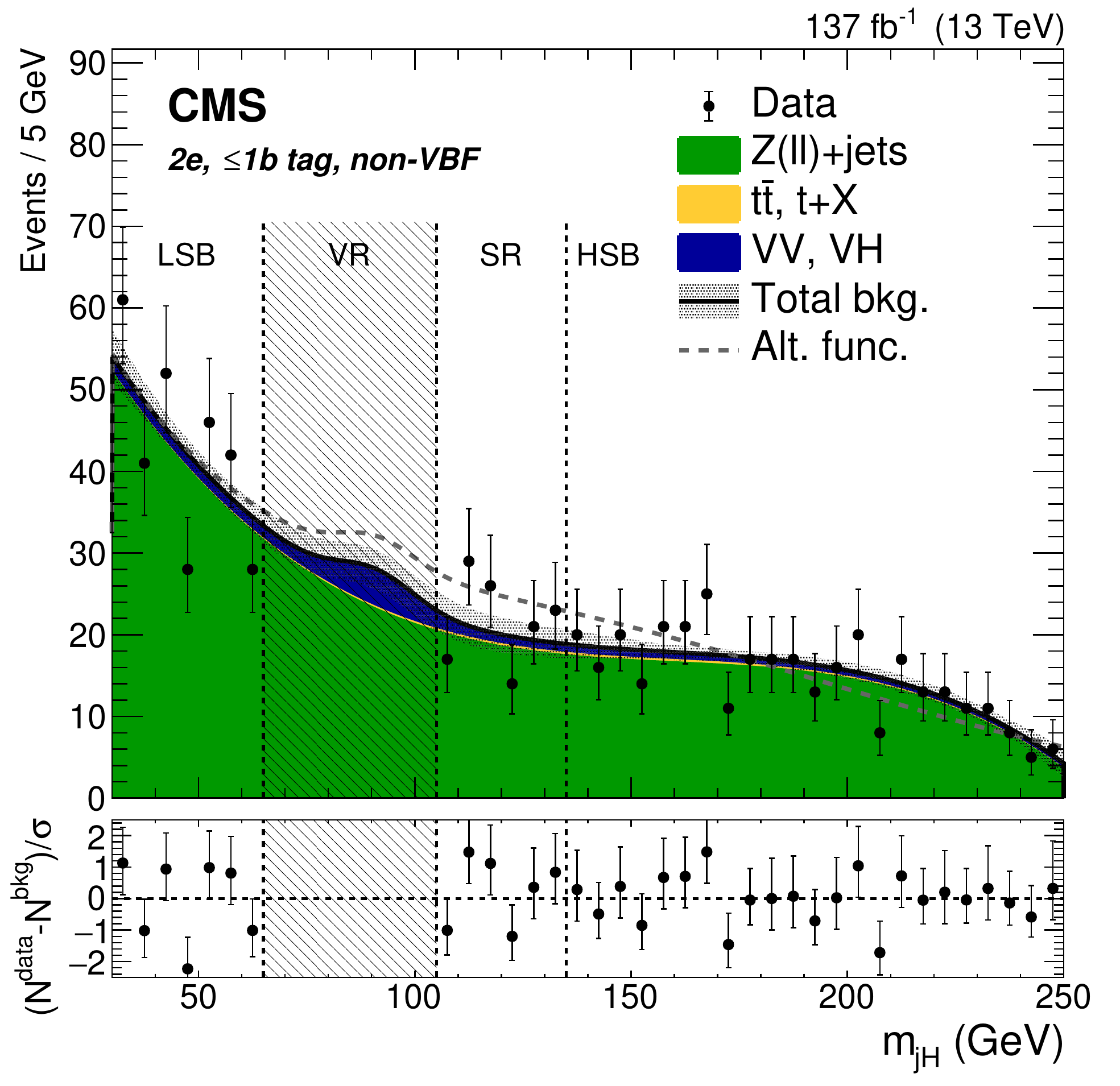}
    \includegraphics[width=.42\textwidth]{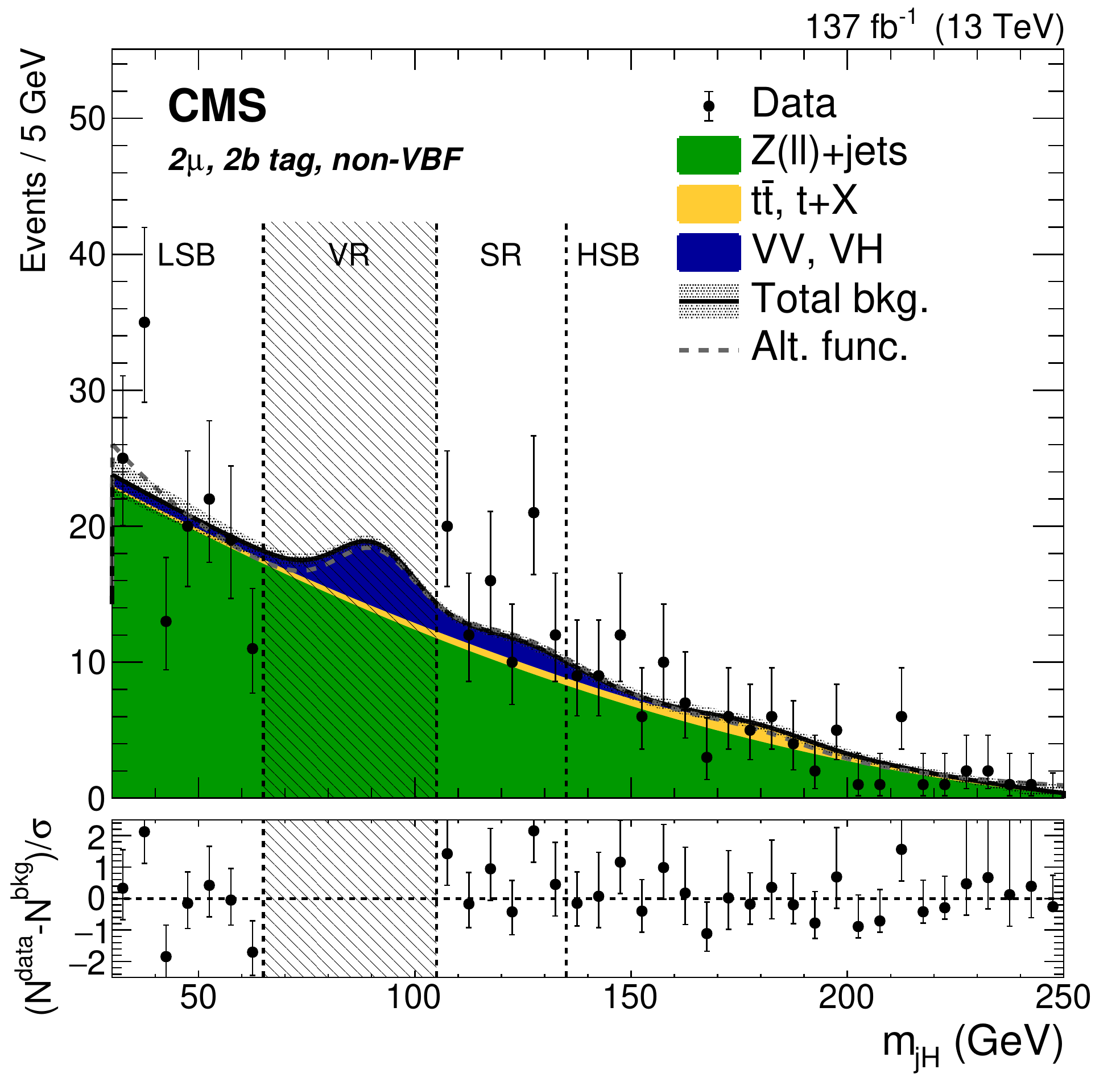}
    \includegraphics[width=.42\textwidth]{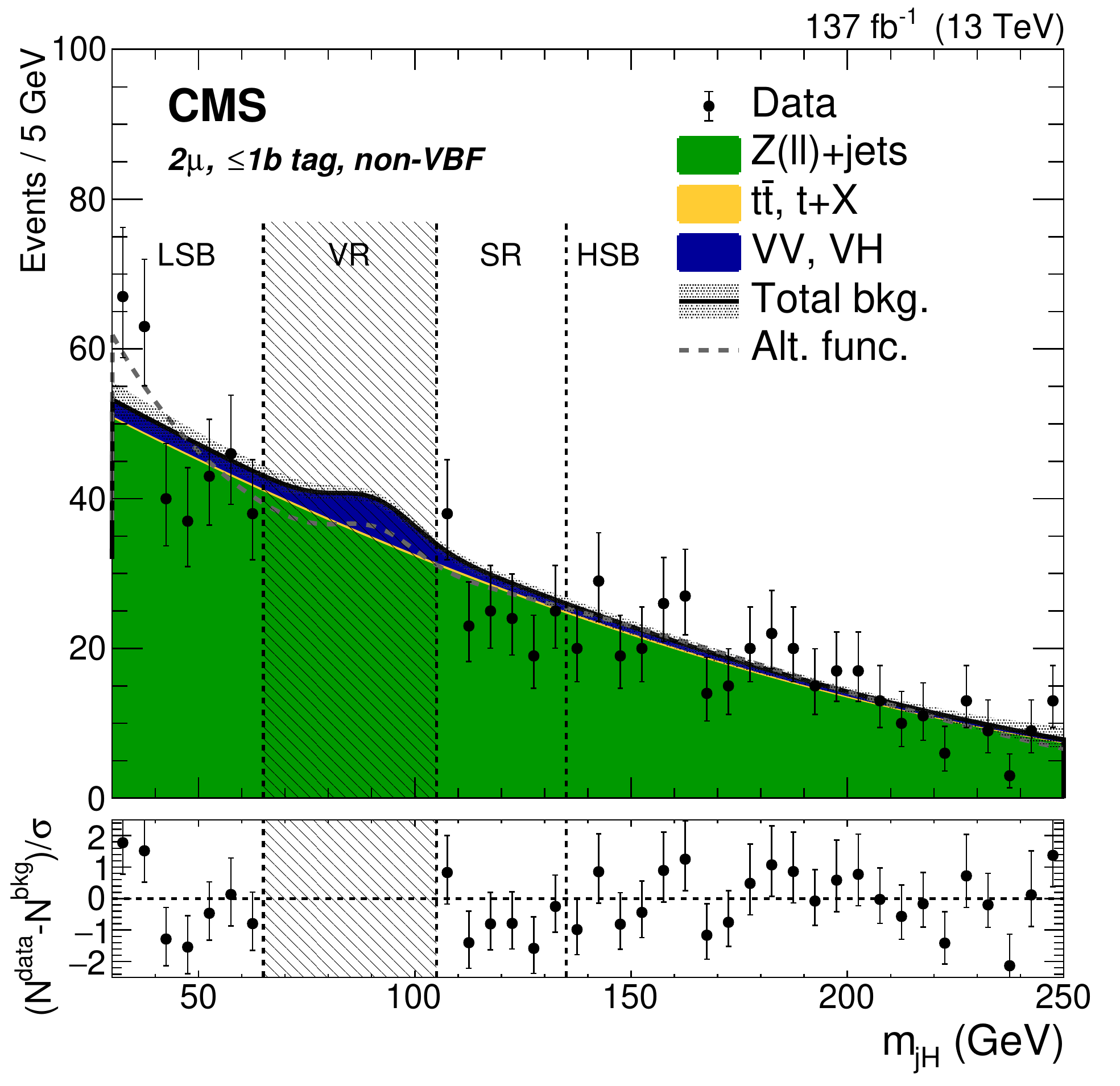}
  \caption{Fit to the \mj distribution in data in the 2\PQb tag (left column) and $\leq$1\PQb tag (right column) non-VBF categories, for $0\ell$ (upper row), 2\Pe (middle row), and 2\PGm (lower row). The shaded bands around the total background estimate represent the uncertainty from the fit to data in the jet mass SBs. The observed data are indicated by black markers. The vertical shaded band indicates the VR region, which is blinded and not used in the fit to avoid potential contamination from \VV resonant signals. The dashed vertical lines separate the LSB, VR, SR, and HSB. The bottom panel shows $(N^{\text{data}}-N^{\text{bkg}})/\sigma$ for each bin, where $\sigma$ is the statistical uncertainty in data. In the $\leq$1\PQb tag, non-VBF categories, \mX or \mtX are required to be larger than 1200\GeV to ensure the smoothness of the background model.}
  \label{fig:XZH_JetMass}
\end{figure*}

\begin{figure*}[!htbp]
  \centering
    \includegraphics[width=.42\textwidth]{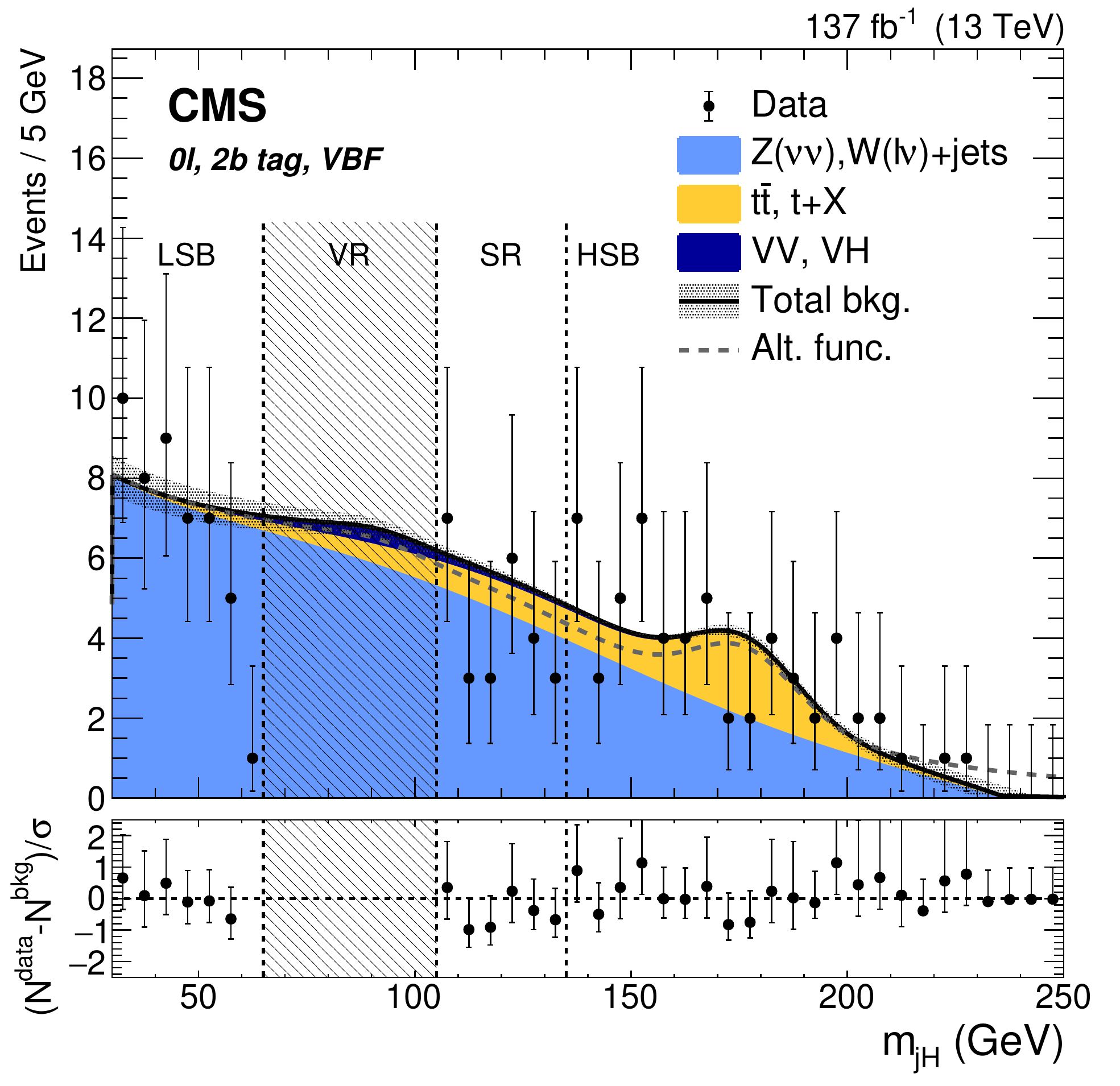}
    \includegraphics[width=.42\textwidth]{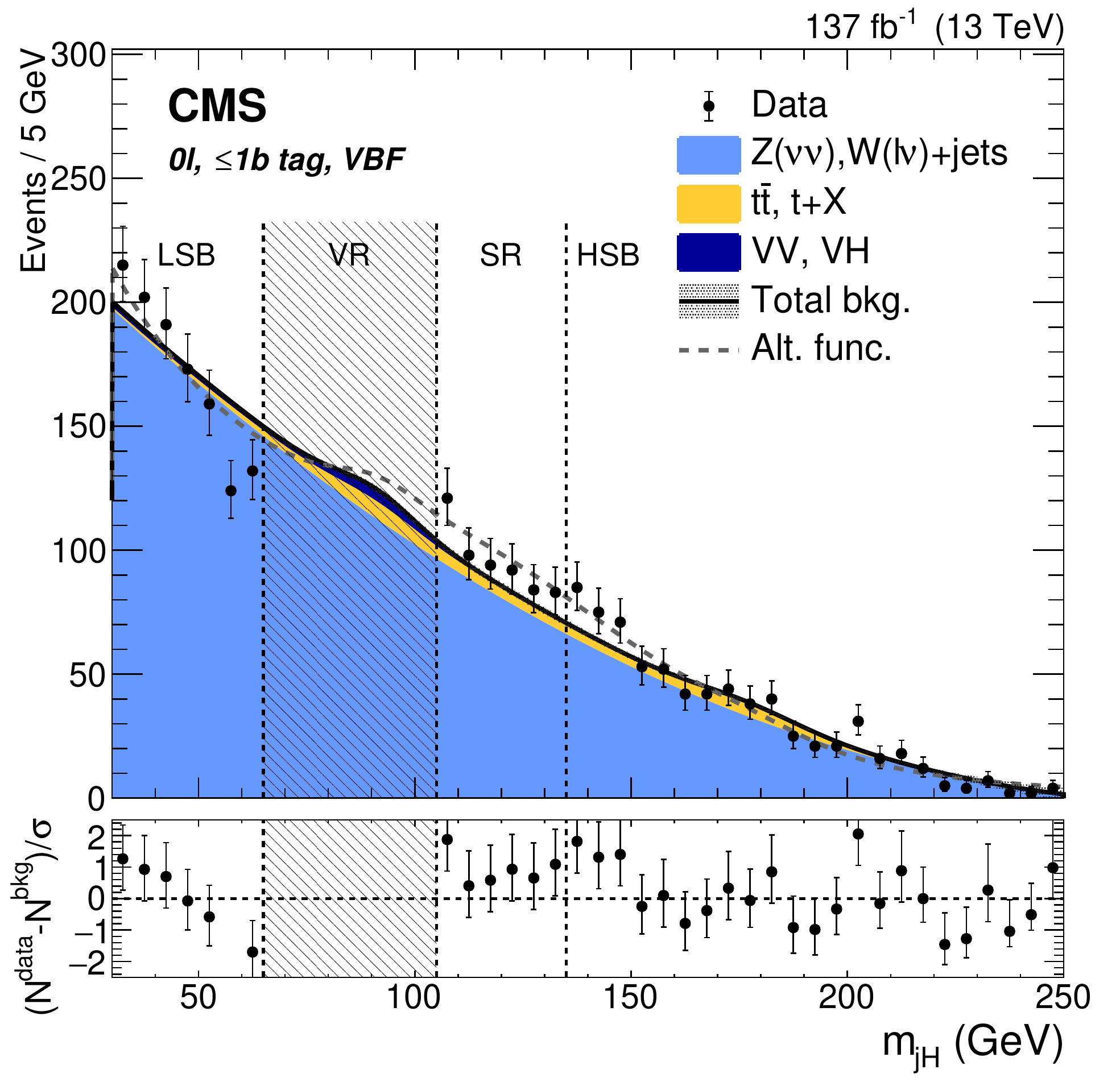}
    \includegraphics[width=.42\textwidth]{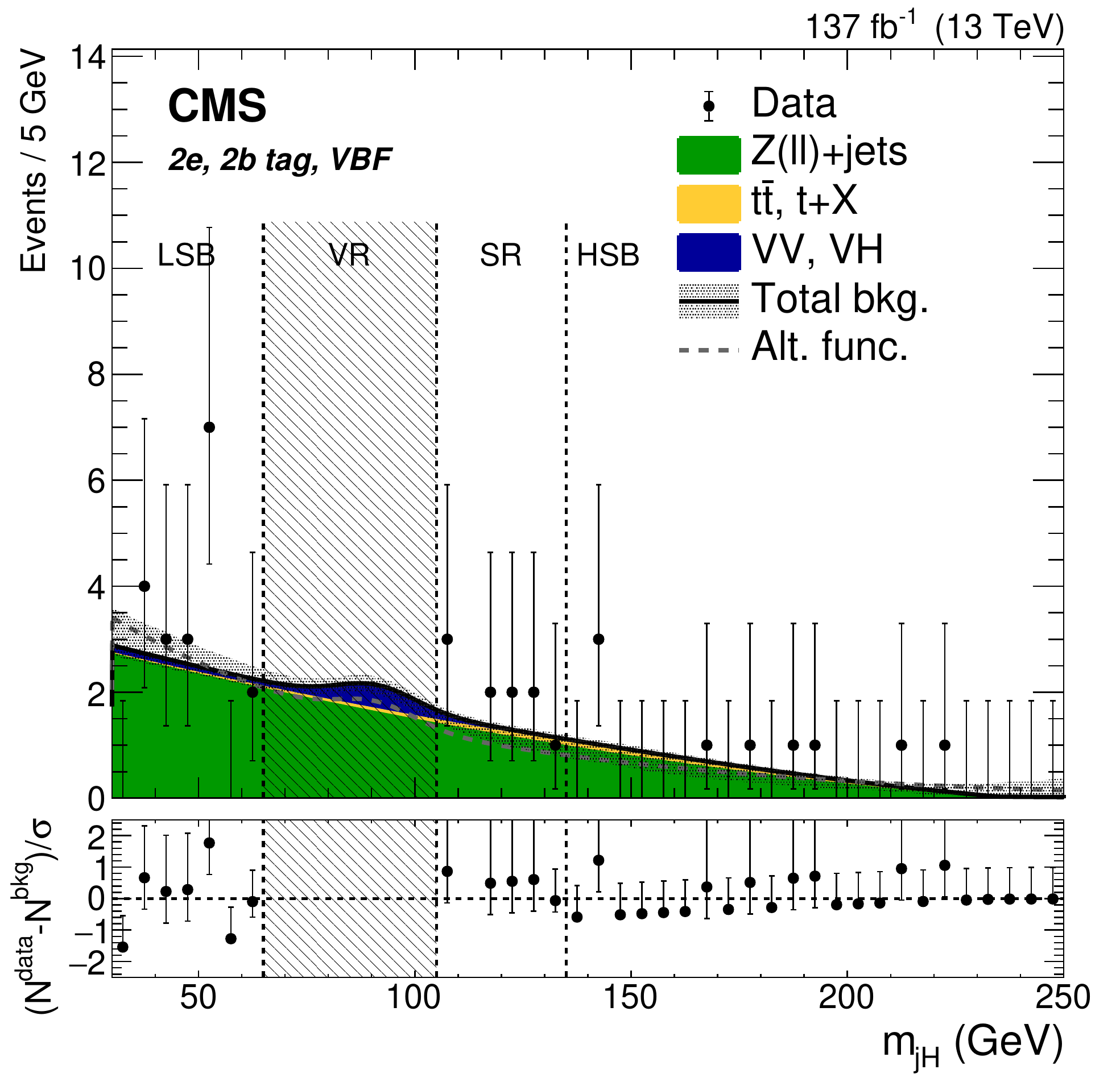}
    \includegraphics[width=.42\textwidth]{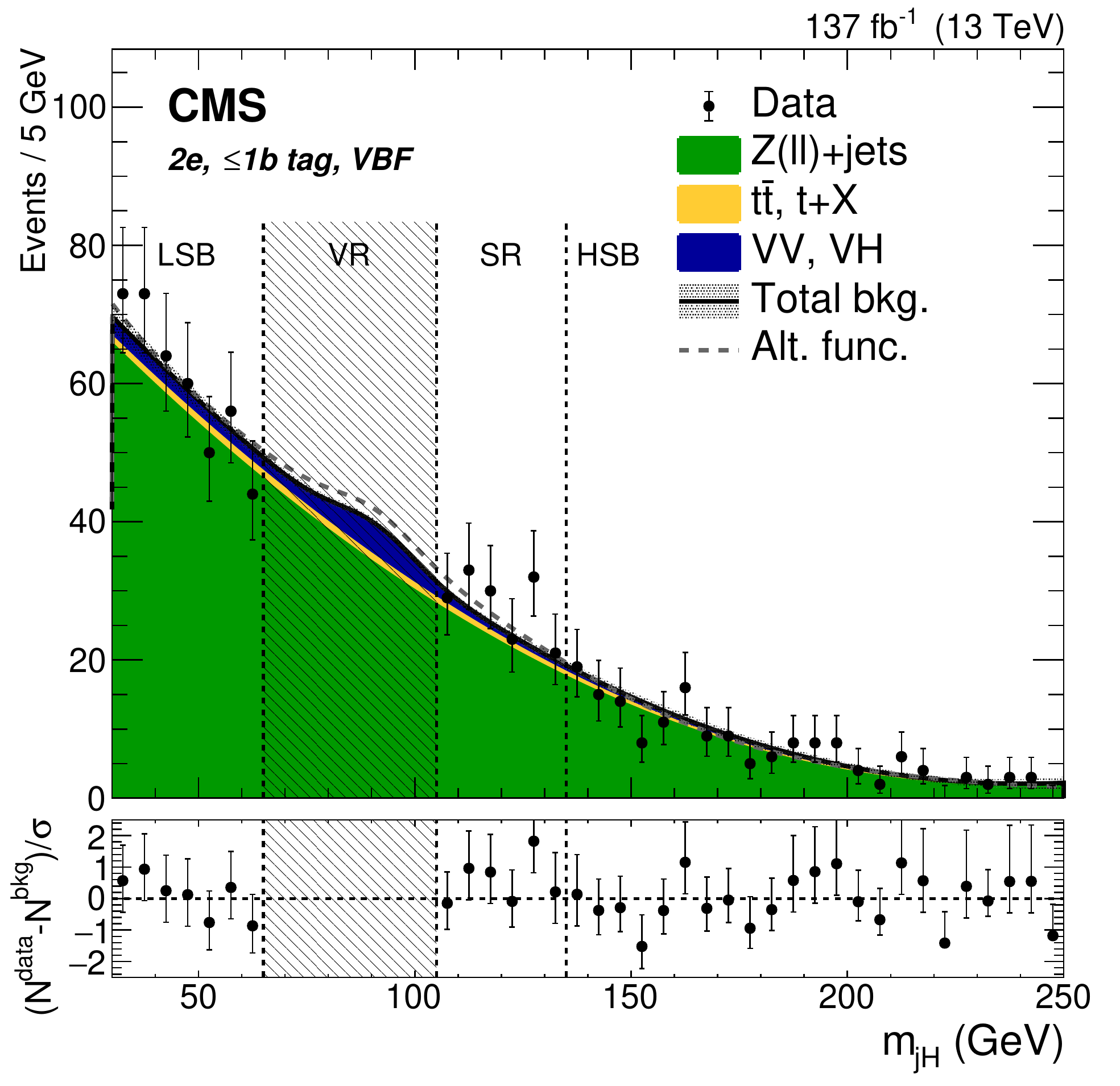}
    \includegraphics[width=.42\textwidth]{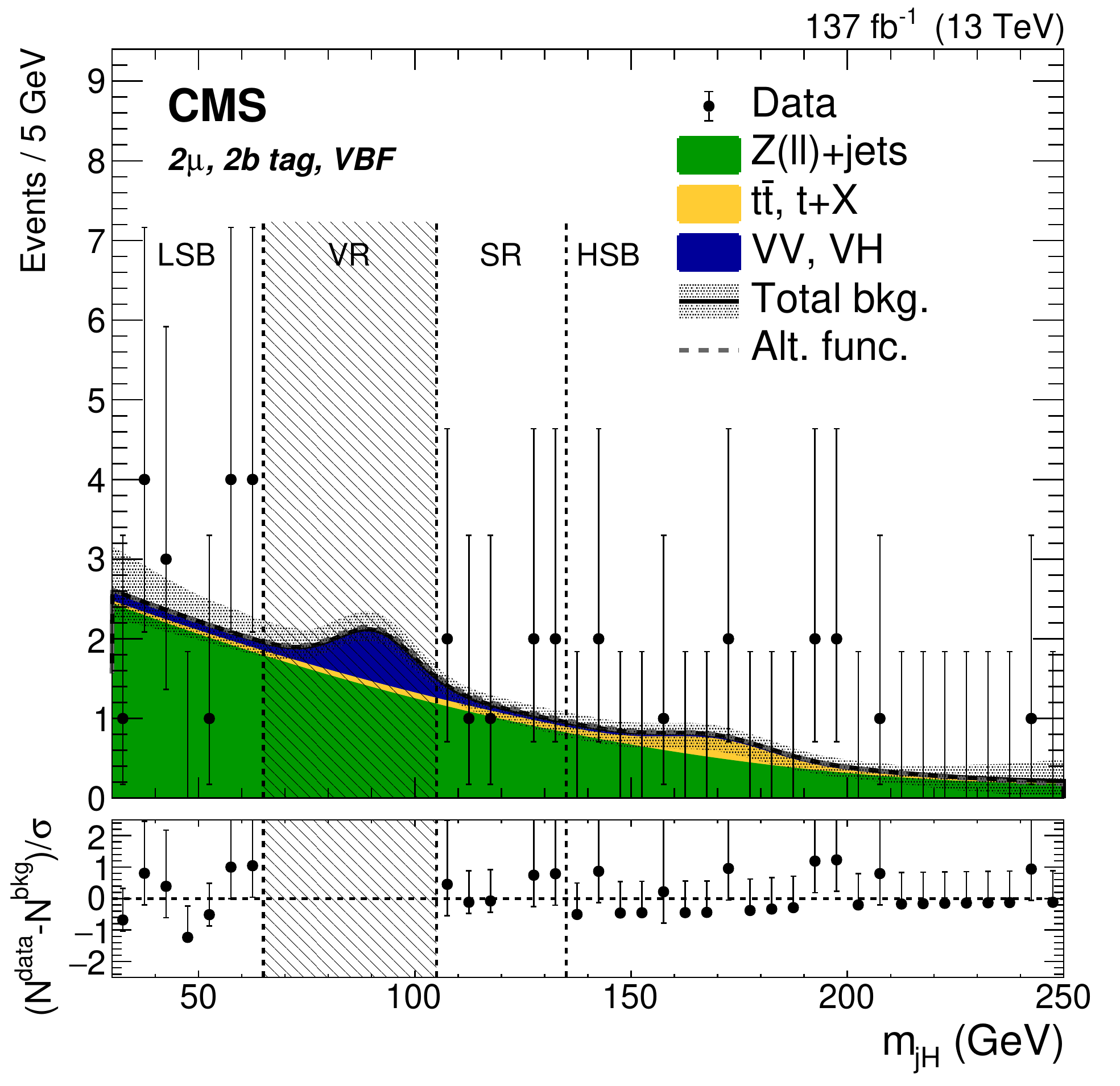}
    \includegraphics[width=.42\textwidth]{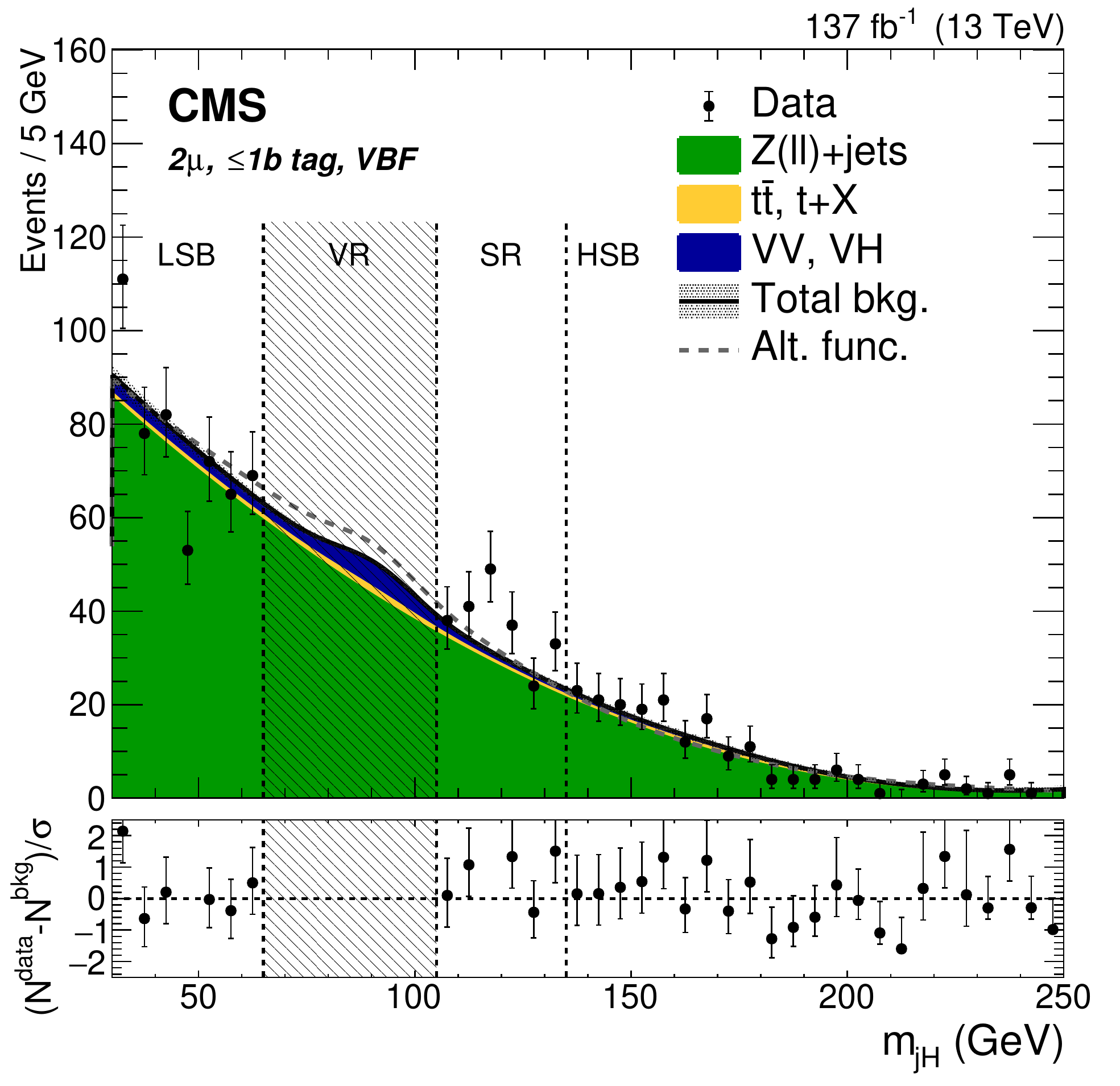}
  \caption{Fit to the \mj distribution in data in the 2\PQb tag (left column) and $\leq$1\PQb tag (right column) VBF categories, for $0\ell$ (upper row), 2\Pe (middle row), and 2\PGm (lower row). The shaded bands around the total background estimate represent the uncertainty from the fit to data in the jet mass SBs. The observed data are indicated by black markers. The observed data are indicated by black markers. The vertical shaded band indicates the VR region, which is blinded and not used in the fit to avoid potential contamination from \VV resonant signals. The dashed vertical lines separate the LSB, VR, SR, and HSB. The bottom panel shows $(N^{\text{data}}-N^{\text{bkg}})/\sigma$ for each bin, where $\sigma$ is the statistical uncertainty in data.}  \label{fig:XZHVBF_JetMass}
\end{figure*}

The background model, composed of the sum of the \Vjets, \ttbar and single top quark, and the \VV and \VH templates is fitted to the SBs of the jet mass distribution in data. The analytical function parameters and the normalization of the top quark and \VV backgrounds are fixed from the fit to simulation, but the shape parameters from the \Vjets background are not. The number of parameters for the fit to data is determined by a Fisher F-test~\cite{Fisher}. The number of expected events is derived from the integral of the fitted model in the SR. The choice of the \Vjets fit function induces a systematic uncertainty, which can be determined by fitting the \Vjets background shape with an alternative function, consisting of the sum of an exponential and a Gaussian function, and considering the difference between the integrals of the two fit models in the SR as a systematic uncertainty. Figures~\ref{fig:XZH_JetMass} and~\ref{fig:XZHVBF_JetMass} show the fits to the jet mass in the different categories. Table~\ref{tab:BkgNorm} summarizes the expected background yield in the SR.

\begin{table*}[!htb]
  \topcaption{The expected and observed numbers of background events in the signal region for all event categories. The \Vjets background uncertainties originate from the variation of the parameters within the fit uncertainties (fit) and the difference between the nominal and alternative function choice for the fit to \mj (alt). The \ttbar and single top quark uncertainties arise from the \mj modeling, the statistical component of the top quark SF uncertainties, and the extrapolation uncertainty from the control region to the SR. The \VV and \VH normalization uncertainties come from the \mj modeling.}
 \newcolumntype{C}{ @{}>{${}}c<{{}$}@{} }
 \centering
 	\cmsTable{
    \begin{tabular}{llc*3{rCl}c}
    	  \hline
      \multicolumn{2}{l}{Non-VBF Category} & \Vjets ($\pm$fit) ($\pm$alt) & \multicolumn{3}{l}{\ttbar, \ST} & \multicolumn{3}{l}{\VV, \VH} & \multicolumn{3}{l}{Bkg. sum} & Observed\\
      \hline
      \multirow{3}{*}{2\PQb tag}
      & $0\ell$ & $374 \pm 34 \pm 20$ & 68 & \pm & 8 & 31& \pm & 10 & 474 & \pm & 42 & 549 \\     
      & $2\Pe$ & $54 \pm 5 \pm 8$ & 3.1 & \pm & 0.4 & 7.9 & \pm & 1.9 & 65 & \pm & 10 & 57 \\
      & $2\PGm$ & $60 \pm 5 \pm 1$ & 3.2 & \pm & 0.6 & 9.1 & \pm & 2.1 & 72 & \pm & 5 & 91 \\[\cmsTabSkip]
      \multirow{3}{*}{$\leq$1\PQb tag}
      & $0\ell$ & $637 \pm 35 \pm 51$ & 7.3 & \pm & 0.9 & 15 & \pm & 4 & 659 & \pm & 61 & 697 \\
      & $2\Pe$ &  $113 \pm 14 \pm 27$ & 1.6 & \pm & 0.2 & 7.2 & \pm & 1.7 & 122 & \pm & 31 & 130 \\
      & $2\PGm$ &  $167 \pm 8 \pm 10$ & 1.8 & \pm & 0.2 & 8.0 & \pm & 1.8 & 177 & \pm & 13 & 154 \\[\cmsTabSkip]
      \hline
      \multicolumn{2}{l}{VBF Category} & \Vjets ($\pm$fit) ($\pm$alt) & \multicolumn{3}{l}{\ttbar, \ST} & \multicolumn{3}{l}{\VV, \VH} & \multicolumn{3}{l}{Bkg. sum} & Observed\\
      \hline
      \multirow{3}{*}{2\PQb tag}
      & $0\ell$ & $28 \pm 3 \pm 3$ & 4.3 & \pm & 2.0 & 0.9 & \pm & 0.6 & 33 & \pm & 5 & 26 \\
      & $2\Pe$ & $7.3 \pm 2.0 \pm 2.0$ & 0.4 & \pm & 0.2 & 0.4 & \pm & 0.1 & 8.1 & \pm & 2.8 & 10 \\
      & $2\PGm$ & $6.0 \pm 1.7 \pm 0.2$ & 0.4 & \pm & 0.2 & 0.5 & \pm & 0.1 & 7.0 & \pm & 1.7 & 8.0 \\[\cmsTabSkip]
      \multirow{3}{*}{$\leq$1\PQb tag}
      & $0\ell$ & $486 \pm 13 \pm 72$ & 25 & \pm & 6 & 6.3 & \pm & 1.5 & 517 & \pm & 73 & 572 \\
      & $2\Pe$ & $137 \pm 7 \pm 7$ & 4.8 & \pm & 1.5 & 6.4 & \pm & 1.5 & 148 & \pm & 10 & 168 \\
      & $2\PGm$ & $171 \pm 8 \pm 6$ & 4.5 & \pm & 1.1 & 7.7 & \pm & 1.8 & 183 & \pm & 10 & 222 \\
      \hline
    \end{tabular}  
    }
   \label{tab:BkgNorm}
\end{table*}

\subsection{Background distribution}
The \mX and \mtX distributions are estimated using the data in the jet mass SBs. An $\alpha$ function is then defined as the ratio of the two functions describing the simulated \mX (or \mtX) shape in the SR and SB region of the \Vjets background:
\begin{equation}
\alpha(m) = \frac{N_{\text{SR}}^{\Vjets}(m)}{N_{\text{SB}}^{\Vjets}(m)},
\end{equation}
where $N$ denotes the function and \emph{m} represents either \mX or \mtX. The functions are normalized to the number of events derived in Section~\ref{subsec:norm} and shown in Table~\ref{tab:BkgNorm}.

\begin{figure*}[!htbp]
  \centering
    \includegraphics[width=.42\textwidth]{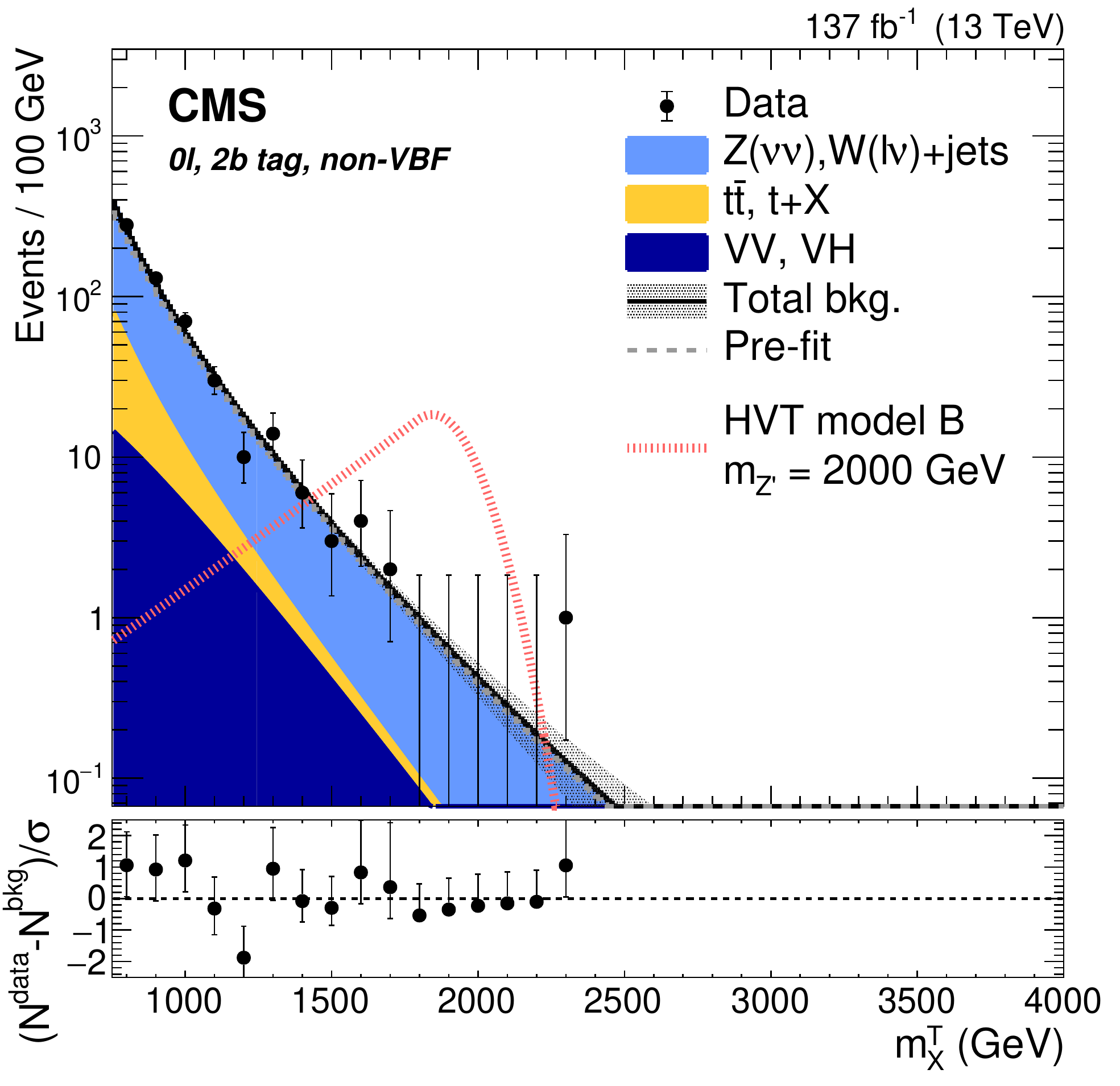}
    \includegraphics[width=.42\textwidth]{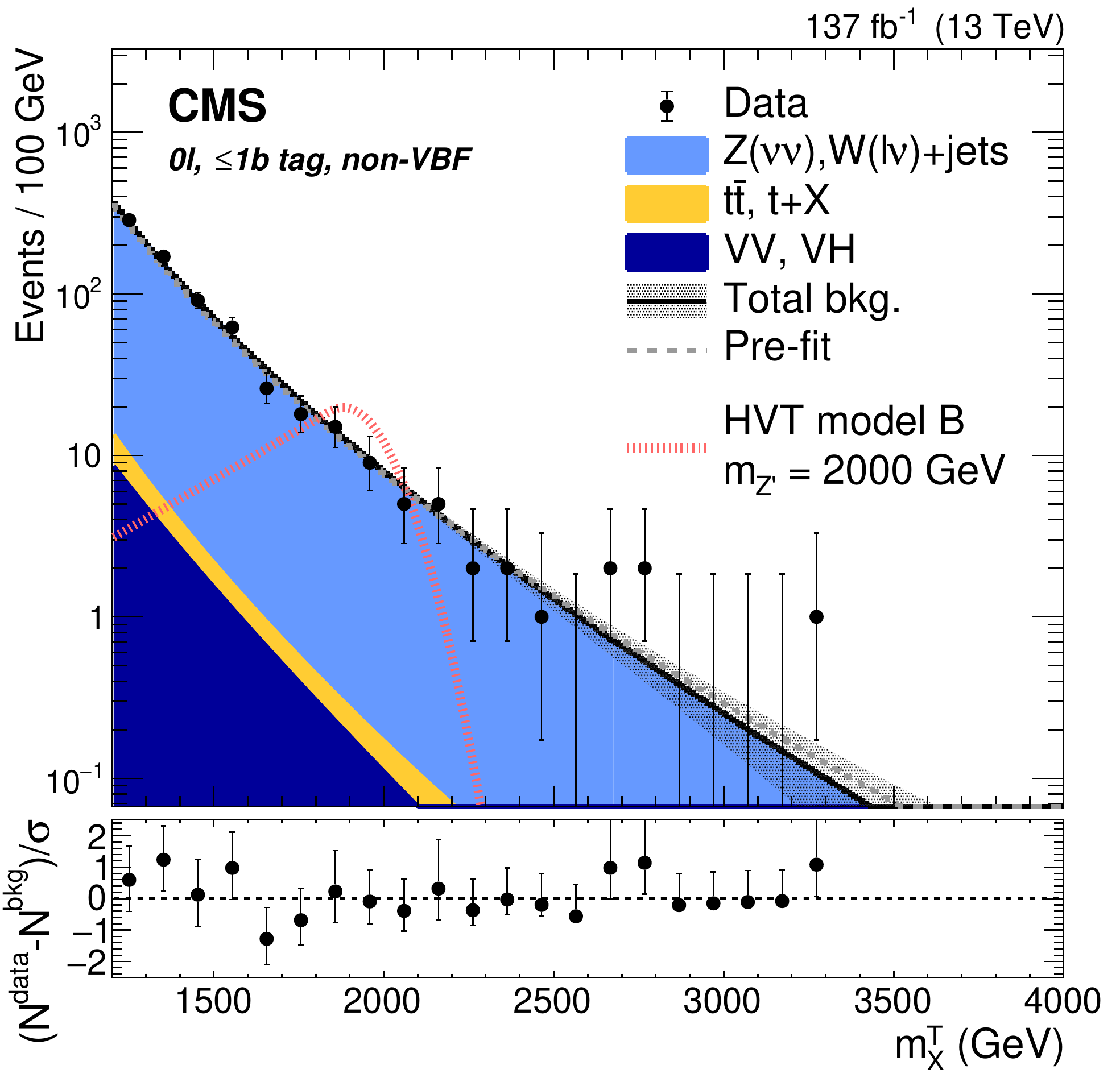}
    \includegraphics[width=.42\textwidth]{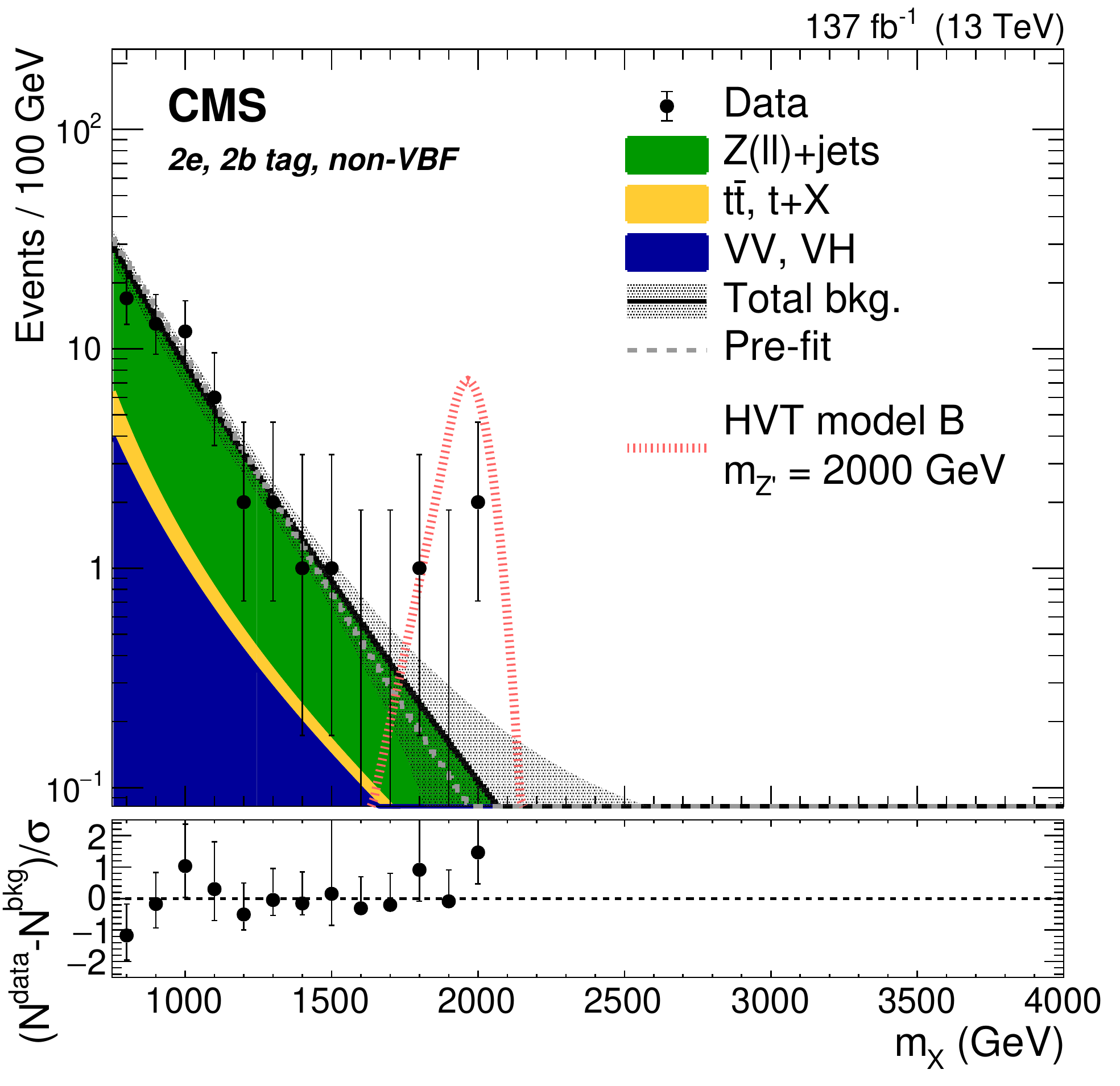}
    \includegraphics[width=.42\textwidth]{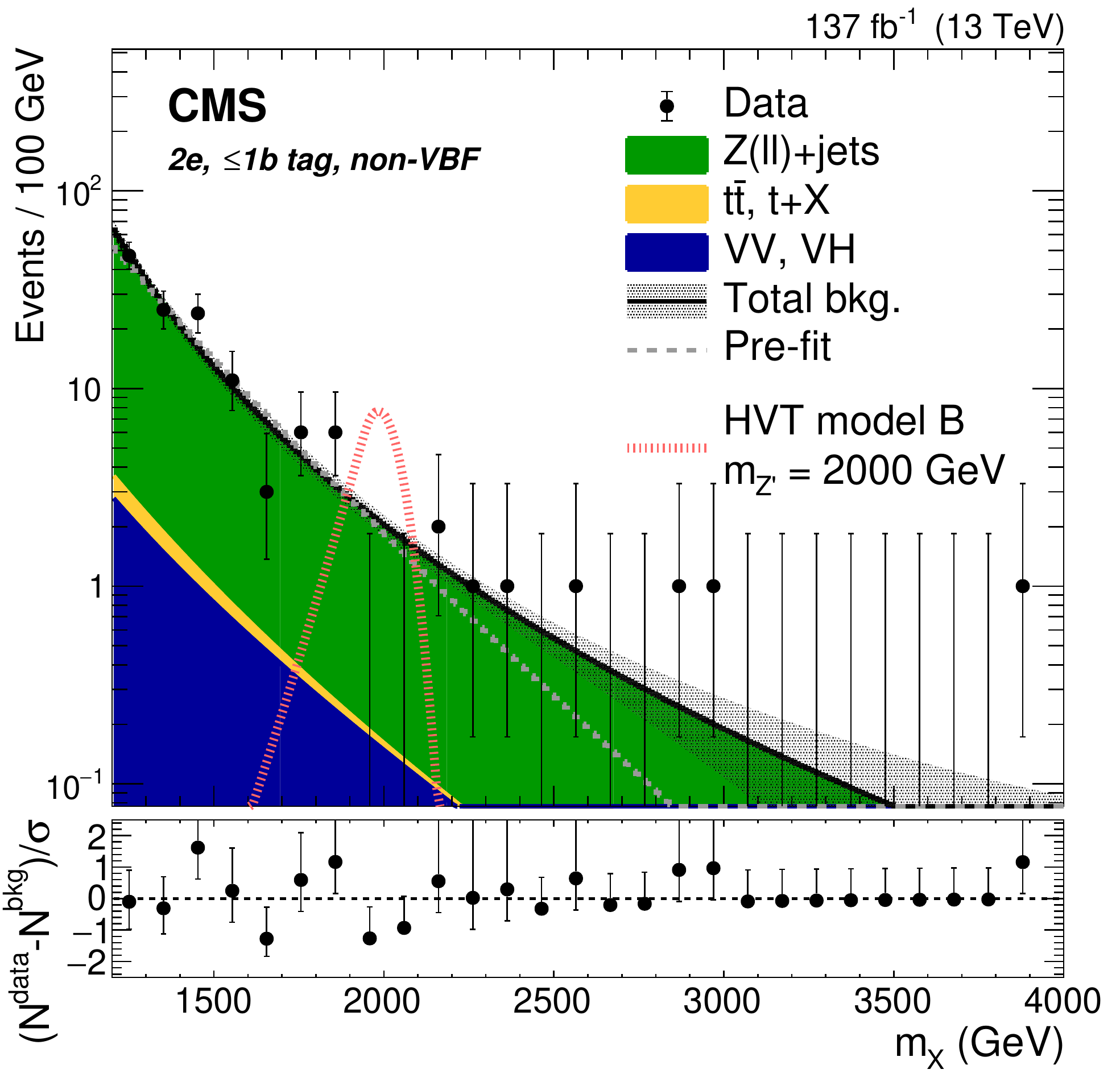}
    \includegraphics[width=.42\textwidth]{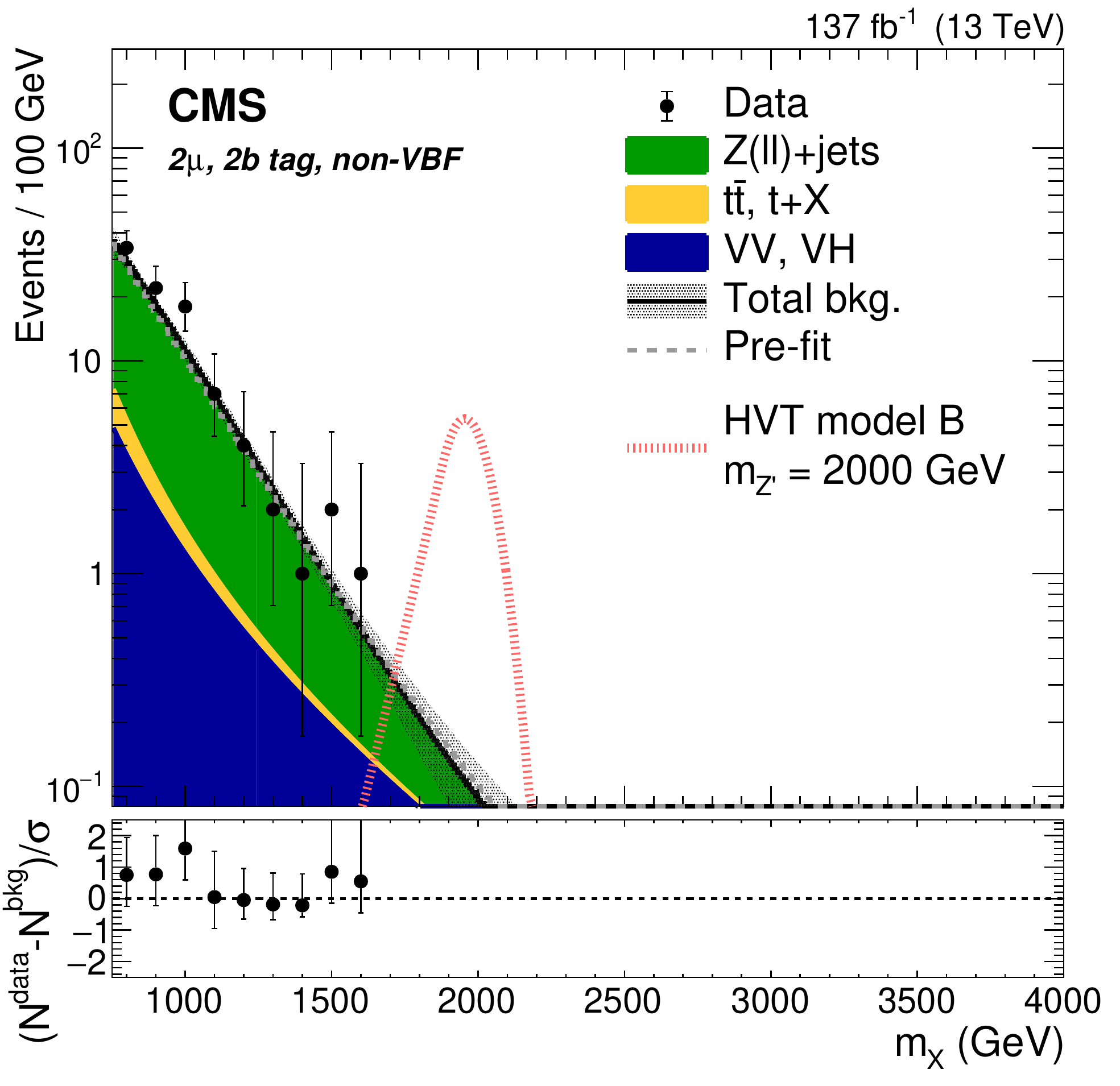}
    \includegraphics[width=.42\textwidth]{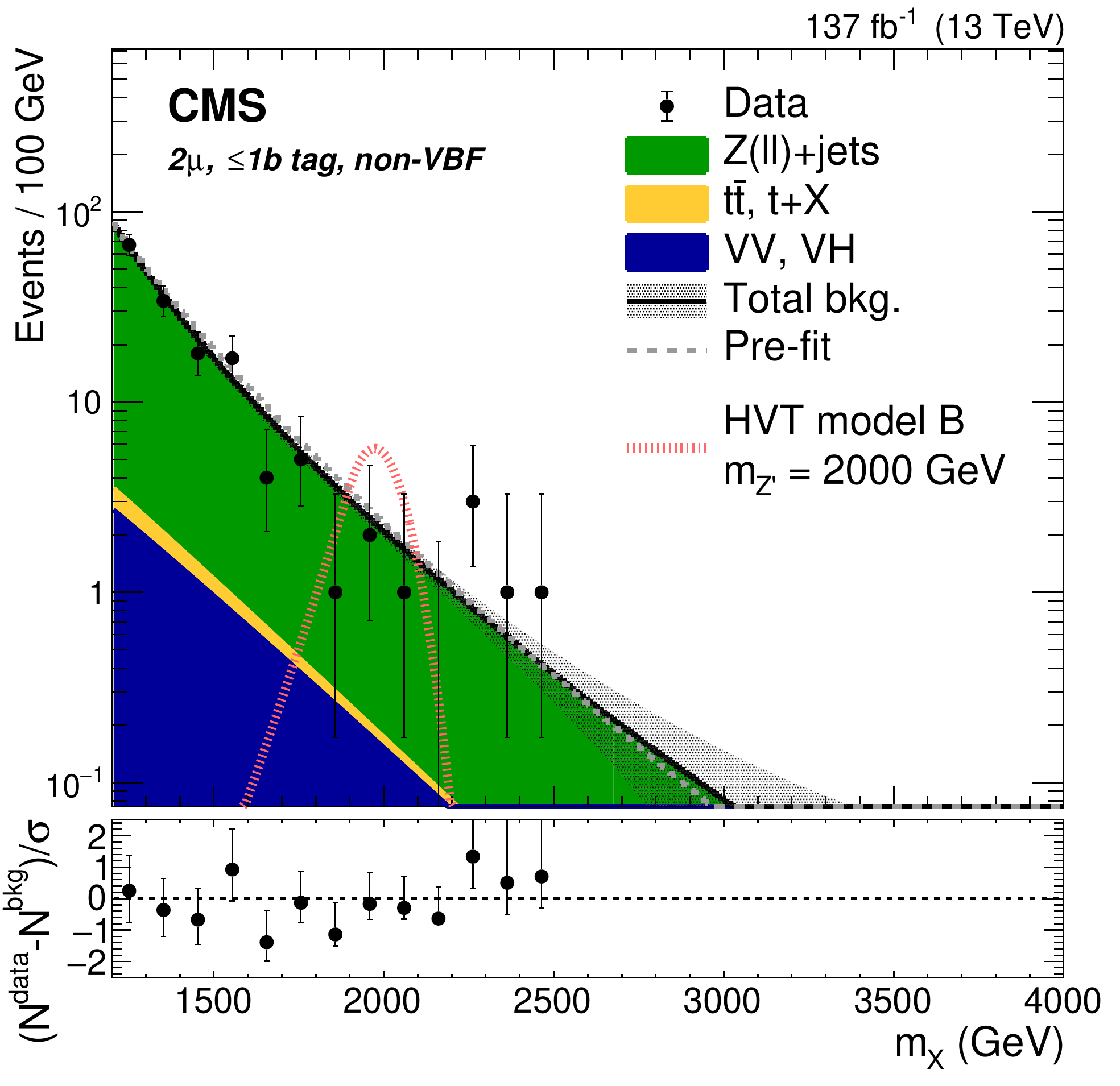}
  \caption{Distributions in data in the 2\PQb tag (left column) and $\leq$1\PQb tag (right column) non-VBF categories, of \mtX for $0\ell$ (upper row), and \mX for 2\Pe (middle row), and 2\PGm (lower row). The distributions are shown up to 4000\GeV, which corresponds to the event with the highest \mX or \mtX observed in the SR. The shaded bands represent the uncertainty from the background estimation. The observed data are represented by black markers, and the potential contribution of a resonance produced in the context of the HVT model B at $\mZpr=2000\GeV$ is shown as a dotted red line. The bottom panel shows $(N^{\text{data}}-N^{\text{bkg}})/\sigma$ for each bin, where $\sigma$ is the statistical uncertainty in data.}
  \label{fig:XZH_BkgSR}
\end{figure*}

\begin{figure*}[!htbp]
  \centering
    \includegraphics[width=.42\textwidth]{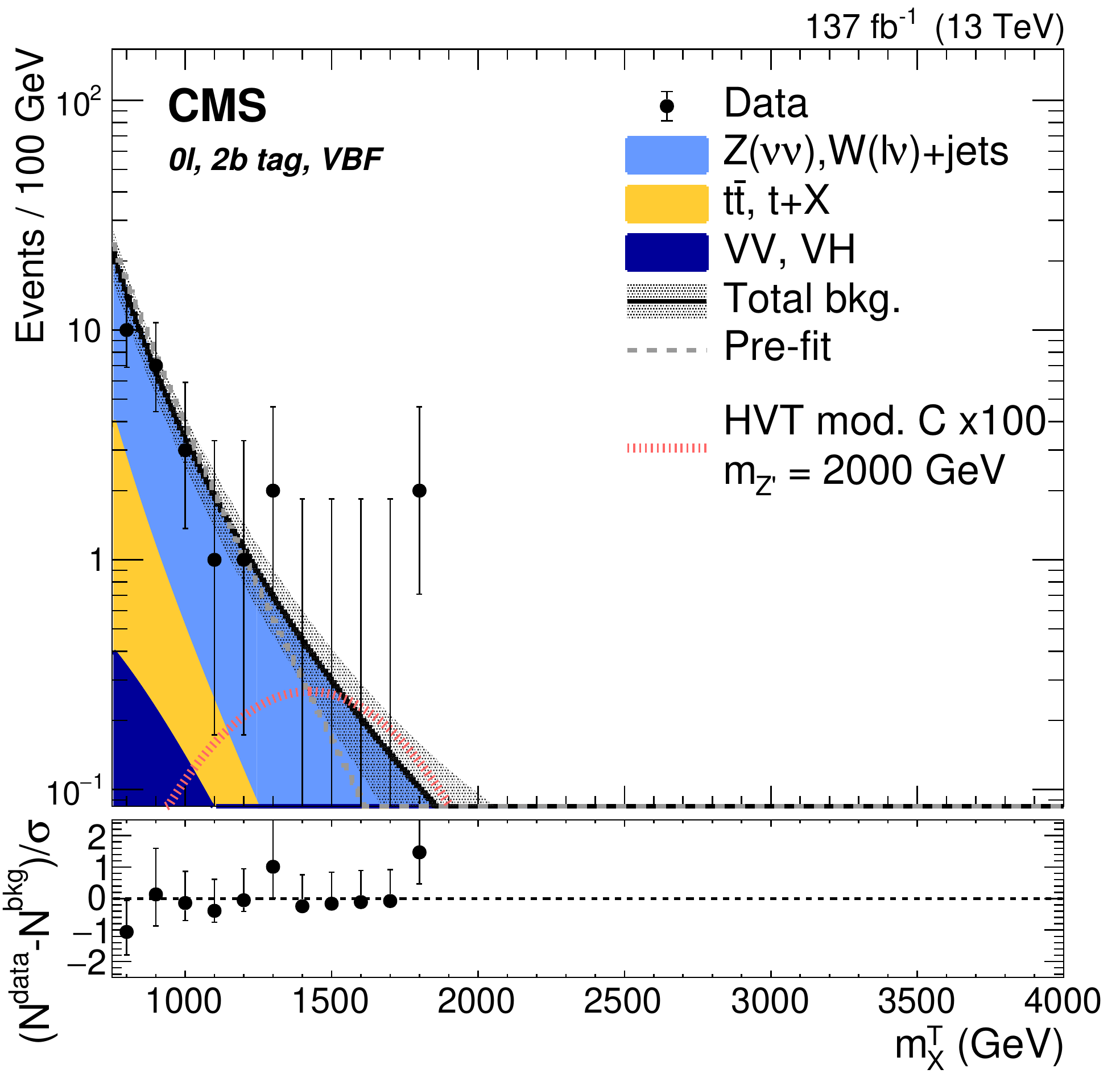}
    \includegraphics[width=.42\textwidth]{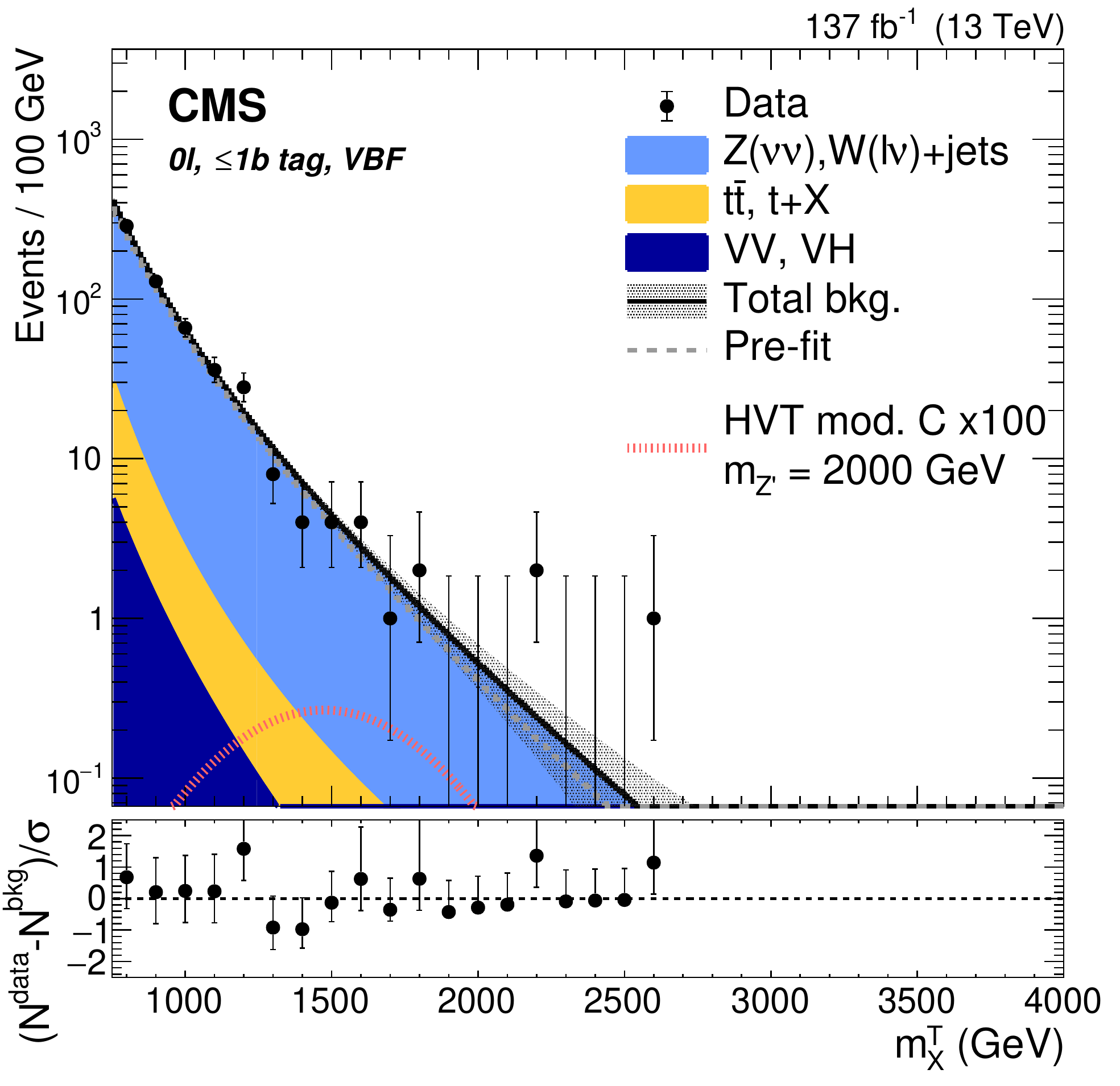}
    \includegraphics[width=.42\textwidth]{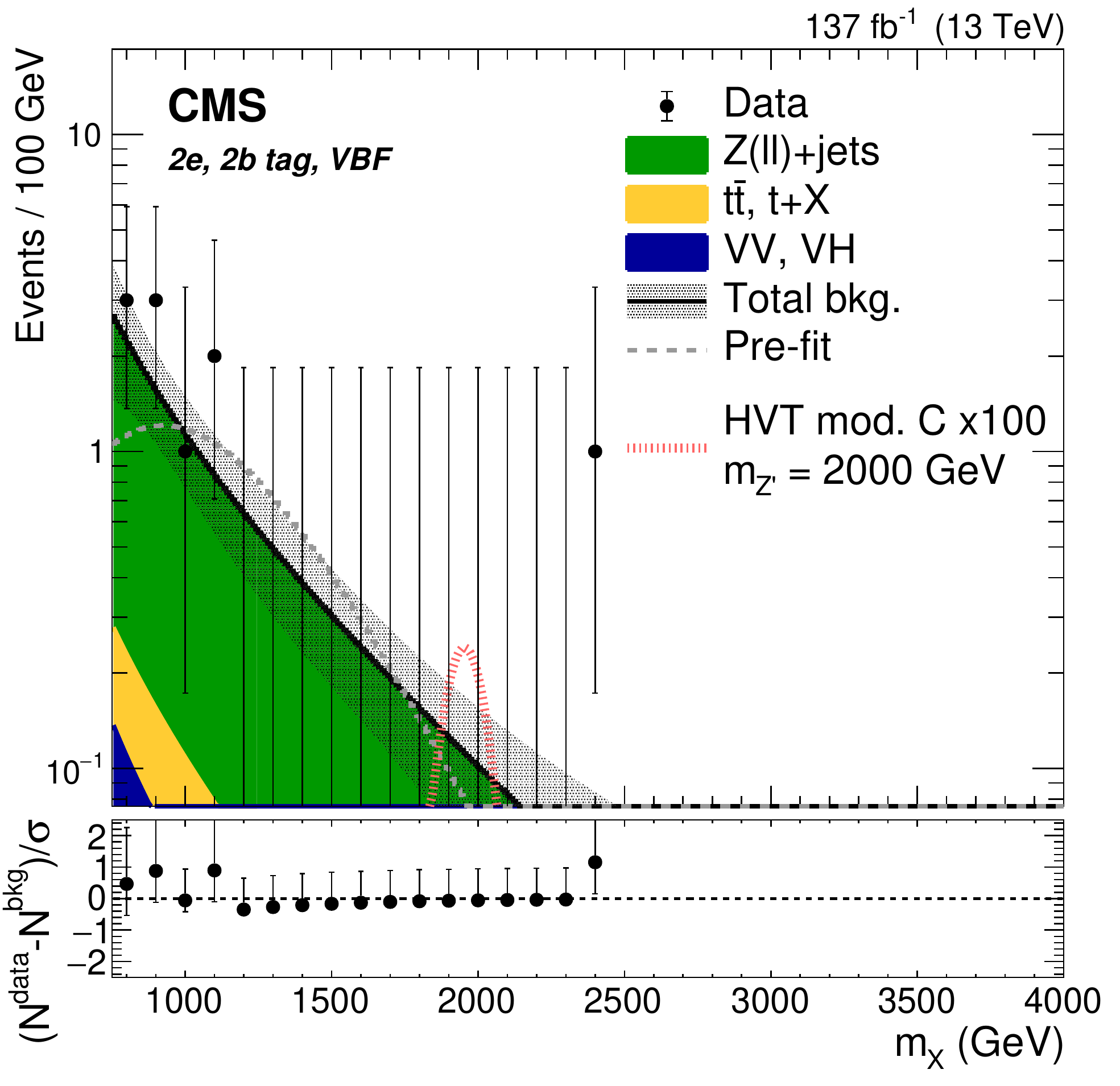}
    \includegraphics[width=.42\textwidth]{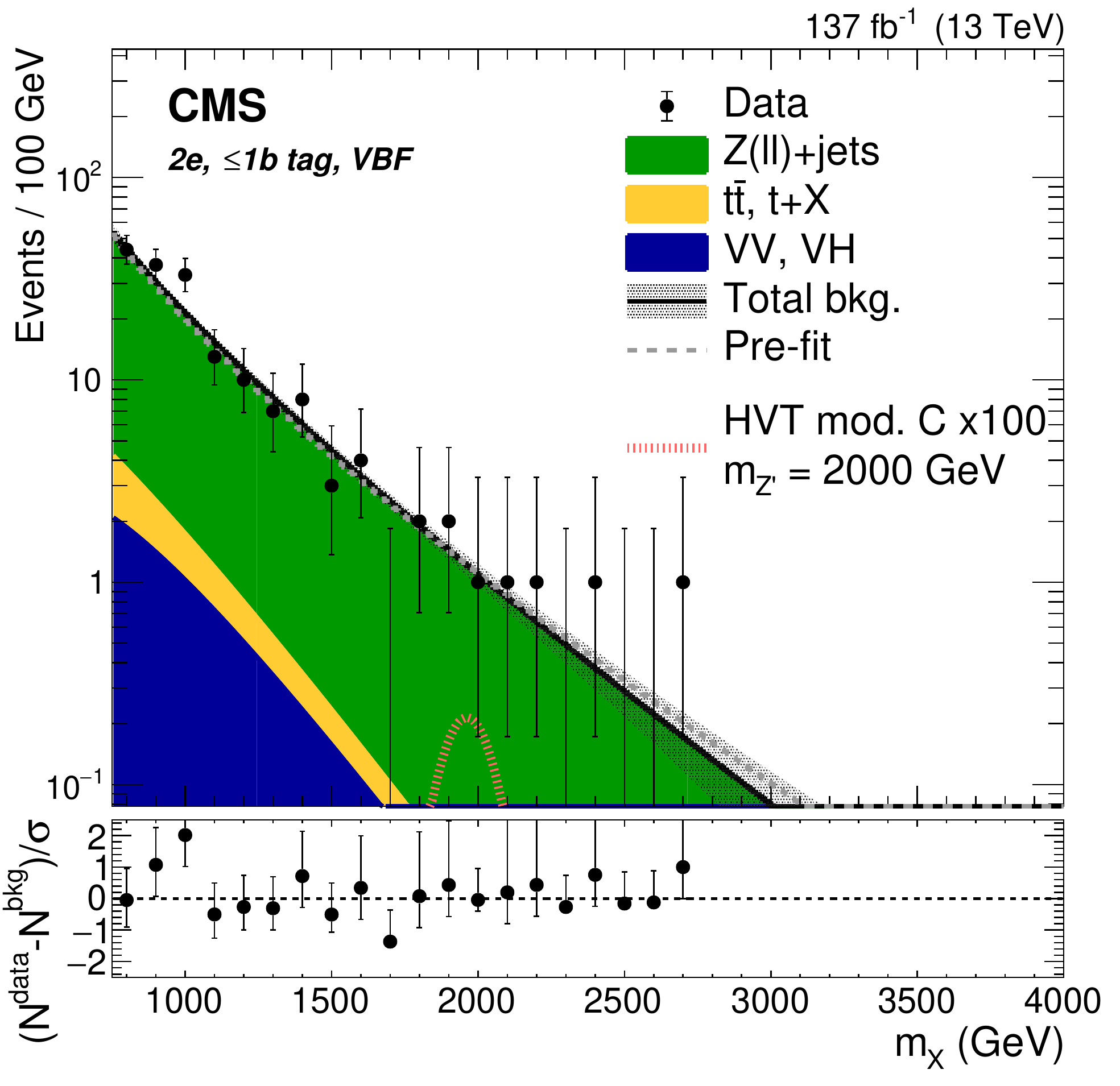}
    \includegraphics[width=.42\textwidth]{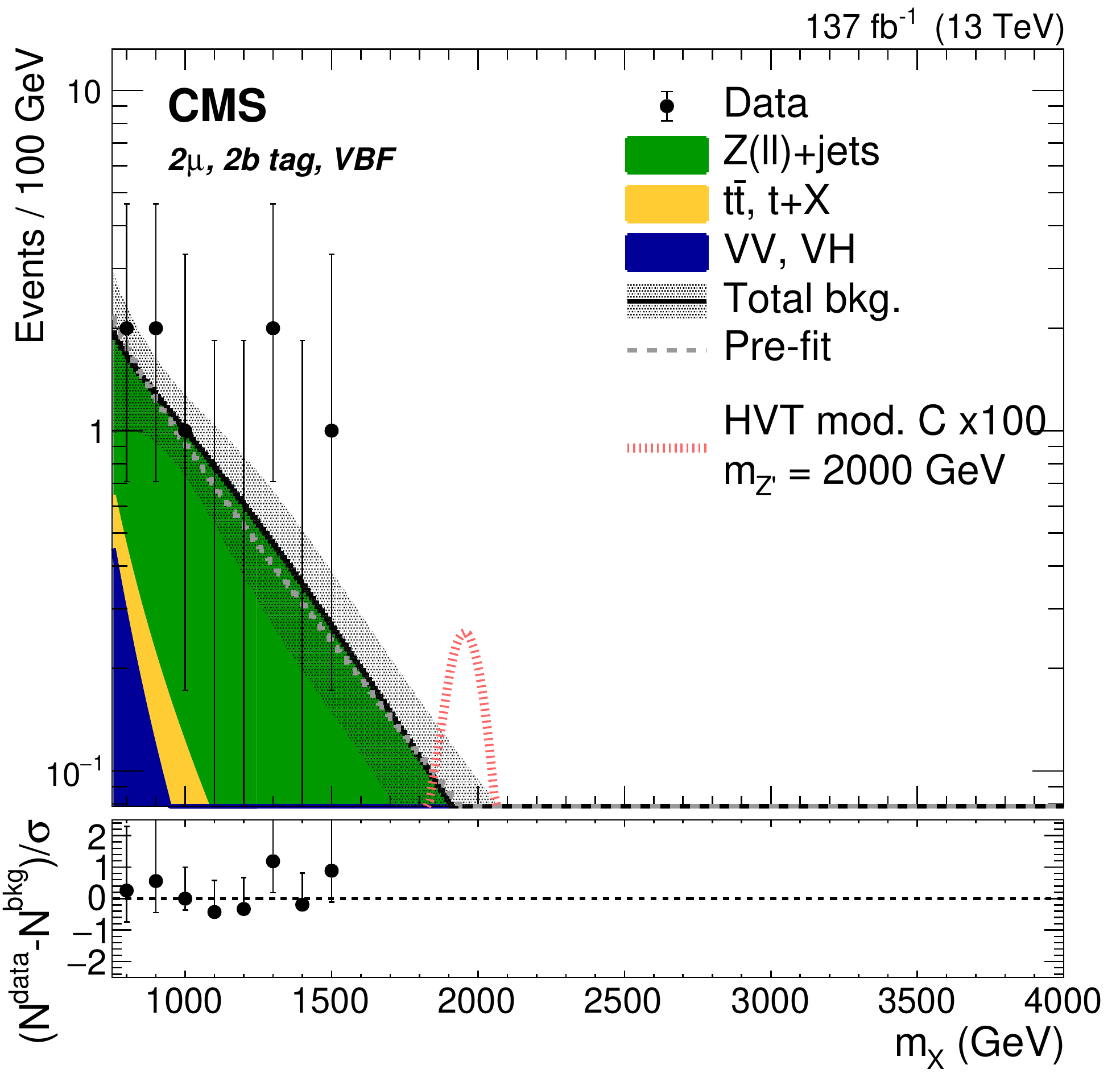}
    \includegraphics[width=.42\textwidth]{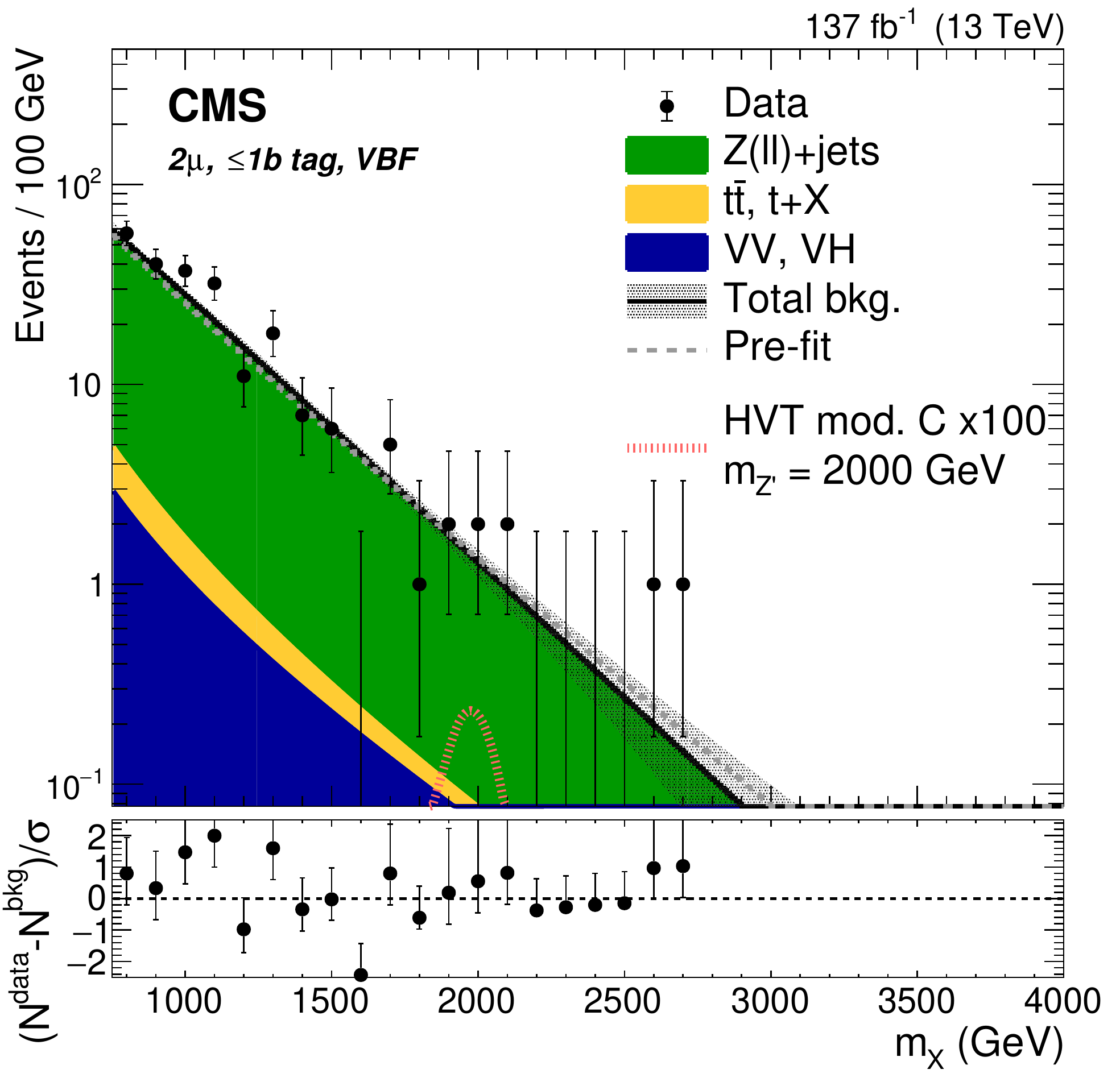}
  \caption{Distributions in data in the 2\PQb tag (left column) and $\leq$1\PQb tag (right column) VBF categories, of \mtX for $0\ell$ (upper row), and \mX for 2\Pe (middle row), and 2\PGm (lower row). The distributions are shown up to 4000\GeV, which corresponds to the event with the highest \mX or \mtX observed in the SR. The shaded bands represent the uncertainty from the background estimation. The observed data are represented by black markers, and the potential contribution of a resonance produced in the context of the HVT model C at $\mZpr=2000\GeV$ is shown as a dotted red line. The bottom panel shows $(N^{\text{data}}-N^{\text{bkg}})/\sigma$ for each bin, where $\sigma$ is the statistical uncertainty in data.}
  \label{fig:XZHVBF_BkgSR}
\end{figure*}

The \Vjets background shape in the SR is thus estimated as the product of $\alpha(m)$ and the shape in the data SBs after subtracting the corresponding top quark and \VV contributions:
\begin{equation}
N_{\text{SR}}^{\Vjets}(m) = \left[ N_{\text{SB}}^{\text{data}}(m) - N_{\text{SB}}^{\text{top}}(m) - N_{\text{SB}}^{\text{VV}}(m) \right] \alpha(m).
\end{equation}
Finally, the expected number of background events in the SR is derived by adding the top quark and \VV contributions to the \Vjets background distribution and taking the \Vjets normalization from the fit to data in the jet mass SBs:
\begin{equation}
N_{\text{SR}}^{\text{bkg}}(m) = N_{\text{SR}}^{\Vjets}(m)  + N_{\text{SR}}^{\text{top}}(m) + N_{\text{SR}}^{\text{VV}}(m).
\end{equation}
The observed data, along with the expected backgrounds, are reported for each category in Figs.~\ref{fig:XZH_BkgSR} and~\ref{fig:XZHVBF_BkgSR}.

The background estimation method is validated by splitting the LSB in two regions: $30<\mj<50\GeV$ and $50<\mj<65\GeV$. The first one is used as a new LSB and the second one as a proxy for the SR. The data yields and distributions are found to be compatible with the expectation in all categories.

\subsection{Signal modeling}

In order to build a template for the signal extraction, the simulated signal mass points are fitted in the SR with the Crystal Ball function~\cite{Oreglia:128793}, which consists of a Gaussian core and a power-law function that describes the low-end tail below a certain threshold. The parameterization for intermediate mass points is determined by linearly interpolating the shape parameters derived by fitting the generated mass points.

\section{Systematic uncertainties}

The systematic uncertainty in the \Vjets background is dominated by the statistical uncertainty of the number of data events in the SBs. The systematic uncertainties in the shape of the \Vjets background are estimated from the covariance matrix of the simultaneous fit of the \mtX and \mX distributions in data in the SBs, and in simulated \Vjets background events in the signal and SB regions. Most of the effect of the uncertainties is correlated among the SB and SR, and cancels out in the $\alpha$ ratio. The \ttbar and \VV background shape uncertainties are propagated from the covariance matrix of the fit to the simulation in the SR. The statistical treatment is consistent with Ref.~\cite{Sirunyan:2019vgt}.

The uncertainty in the top quark background normalization originates from a limited event count in data and simulated event samples in the control regions, and from the variations on the requirements of lepton selection, \PQb tagging SFs, and the VBF selection used to select events in the control region. The uncertainties are reported in Table~\ref{tab:TopCR}. The uncertainties in the trigger, identification, and isolation efficiencies of leptons affect the normalization and shape of the simulated signal and diboson background. The uncertainties are evaluated by moving the SFs, derived as the efficiency in data over the efficiency in simulation, up and down by one standard deviation, and amount to 1--7\%. 

The lepton scale and resolution affect both shape and normalization of the signal, leading to an uncertainty of 1--3\%. The uncertainty from the effect of the \ptmiss scale and resolution on the normalization of the signal and \VV,\VH background is 1\%. The jet energy scale and resolution uncertainties amount to a 1\% systematic uncertainty in the normalization and a shape variation in the distribution of the signal and diboson background events. The uncertainty in the jet mass scale (resolution) adds a contribution of 0.6 (9.0)\%) to the uncertainty in the signal and the diboson background normalization. The jet mass scale and resolution depend on the choice of the parton shower model, which affects the Higgs boson tagging and leads to an additional uncertainty of 6\% in the signal normalization. The uncertainty was evaluated by using \HERWIGpp 2.7.1~\cite{B_hr_2008} as an alternative showering algorithm. The impact of the \PQb tagging systematic uncertainty in the signal efficiency depends on the mass of the resonance and has a range of 4--15\% for the 2\PQb tag categories and 1--6\% for the $\leq$1\PQb tag categories. The uncertainty is treated as anti-correlated between the two \PQb tag categories.

The event yields and acceptances are affected by the choice of the parton distribution functions (PDFs) and the QCD factorization and renormalization scale uncertainties. The effects of the PDF choice on the acceptance and normalization of the \PZpr signal are derived according to the PDF4LHC recommendations~\cite{Butterworth_2016} and amount to 0.5\% in the acceptance and 8--30\% in the normalization of the signal, 0.2\% in the acceptance and 4.7\% in the normalization of the \VV,\VH background, and 0.1\% in the acceptance and 0.1\% in the normalization of the \ttbar background. The factorization and renormalization scale uncertainties are 3--15\%, depending on the resonance mass for the signal, 18.9\% for the \VV,\VH background, and 1\% for the extrapolation of the top quark SFs to the SR. 

The darkening of ECAL crystals, due to radiation damage, leads to a gradual timing shift, which was not properly propagated to the level 1 trigger for 2016 and 2017~\cite{2006.10165}. This effect is accounted for by adding a 1\% systematic uncertainty in the signal normalization. Additional systematic uncertainties come from estimations of the pileup contribution and the integrated luminosity~\cite{CMS-PAS-LUM-17-001,CMS-PAS-LUM-17-004,CMS-PAS-LUM-18-002}. A list of all systematic uncertainties is given in Table~\ref{tab:Sys}.

\begin{table*}[!htb]
  \centering
  \topcaption{Summary of systematic uncertainties for the background and signal samples. The entries labeled with $\dagger$ are also propagated to the shapes of the distributions. Uncertainties marked with $\ddagger$ impact the signal cross section. Uncertainties in the same line are treated as correlated. All uncertainties except for in the integrated luminosity are considered correlated across the three years of data taking.}
  \label{tab:Sys}
  \cmsTable{
  \begin{tabular}{lcccc}
    \hline
                                    & \Vjets & \ttbar, \ST & \VV, \VH & Signal \\
    \hline
    Bkg. normalization              & 6--40\% & \NA & \NA & \NA \\
    Top quark background SFs        & \NA & 0.4--9.5\% & \NA & \NA \\
    Electron id., isolation  & \NA & \NA & \multicolumn{2}{c}{3.6\%} \\
    Muon id., isolation   & \NA & \NA & \multicolumn{2}{c}{4.9\%} \\
    Electron trigger                & \NA & \NA & \multicolumn{2}{c}{0.9\%} \\
    Muon trigger                	  & \NA & \NA & \multicolumn{2}{c}{7\%} \\
    Lepton scale and resolution $\dagger$ & \NA & \NA & \NA & 1--3\% \\
    \ptmiss scale and resolution             & \NA & \NA & \multicolumn{2}{c}{1\%} \\
    Jet energy scale $\dagger$      & \NA & \NA & 1.0\% & 1.0\% \\
    Jet energy resolution $\dagger$ & \NA & \NA & 0.1\% & 0.1\% \\
    Jet mass scale                  & \NA & \NA & 0.6\% & 0.6\% \\
    Jet mass resolution             & \NA & \NA & 9.0\% & 9.0\% \\
    Higgs boson tagging             & \NA & \NA & \NA & 6\%\\
    \PQb tagging                       & \NA & 1.4\% ($0\ell$) & 0.6\% ($\leq$1\PQb), 6.5\% (2\PQb) & 1-6\% ($\leq$1\PQb), 4-15\% (2\PQb) \\
    PDF, normalization                       & \NA & 0.1\% & 4.7\% & 8--30\% $\ddagger$ \\
    PDF, acceptance                  & \NA & 0.1\% & 0.2\% & 0.5\% \\
    QCD renormalization and		   & \multirow{2}{*}{\NA} & \multirow{2}{*}{\NA} & \multirow{2}{*}{18.9\%} & \multirow{2}{*}{3--15\% $\ddagger$} \\
    \hspace{1em}factorization scales\\
    Factorization and renorm.        & \multirow{2}{*}{\NA} & \multirow{2}{*}{1\%} & \multirow{2}{*}{\NA} & \multirow{2}{*}{\NA} \\
    \hspace{1em}scales extrapolation\\
    Level 1 trigger                & \NA & \NA & \NA & 1\% \\
    Pileup                         & \NA & \NA & 0.1\% & 0.1\% \\
    Integrated luminosity                      & \NA & \NA & 1.8\% & 1.8\% \\
    \hline
  \end{tabular}
  }
\end{table*}

\section{Results}

Results are obtained from a combined profile likelihood fit to the unbinned \mtX and \mX distributions of signal and background, shown in Figs.~\ref{fig:XZH_BkgSR} and~\ref{fig:XZHVBF_BkgSR}. Systematic uncertainties are treated as nuisance parameters and are profiled in the statistical interpretation~\cite{bib:CLS2,bib:CLS1,CMS-NOTE-2011-005}. The uncertainties in the signal normalization that are derived from the signal cross section are not profiled in the likelihood, and are reported separately as the uncertainty band of the theoretical cross section. The statistical methods, including the treatment of the nuisance parameters, are described in more detail in Ref.~\cite{Sirunyan:2019vgt}.

The background-only hypothesis is tested against a hypothesis also considering \ZprtoZH signal in all categories. A modified frequentist method is used to determine 95\% confidence level (CL) upper limits on the product of cross section and branching fraction as a function of \mX, in which the distribution of the profile likelihood test statistic is derived using an asymptotic approximation~\cite{bib:Asymptotic}.

\begin{figure*}[!htb]
  \centering
  \includegraphics[width=.49\textwidth]{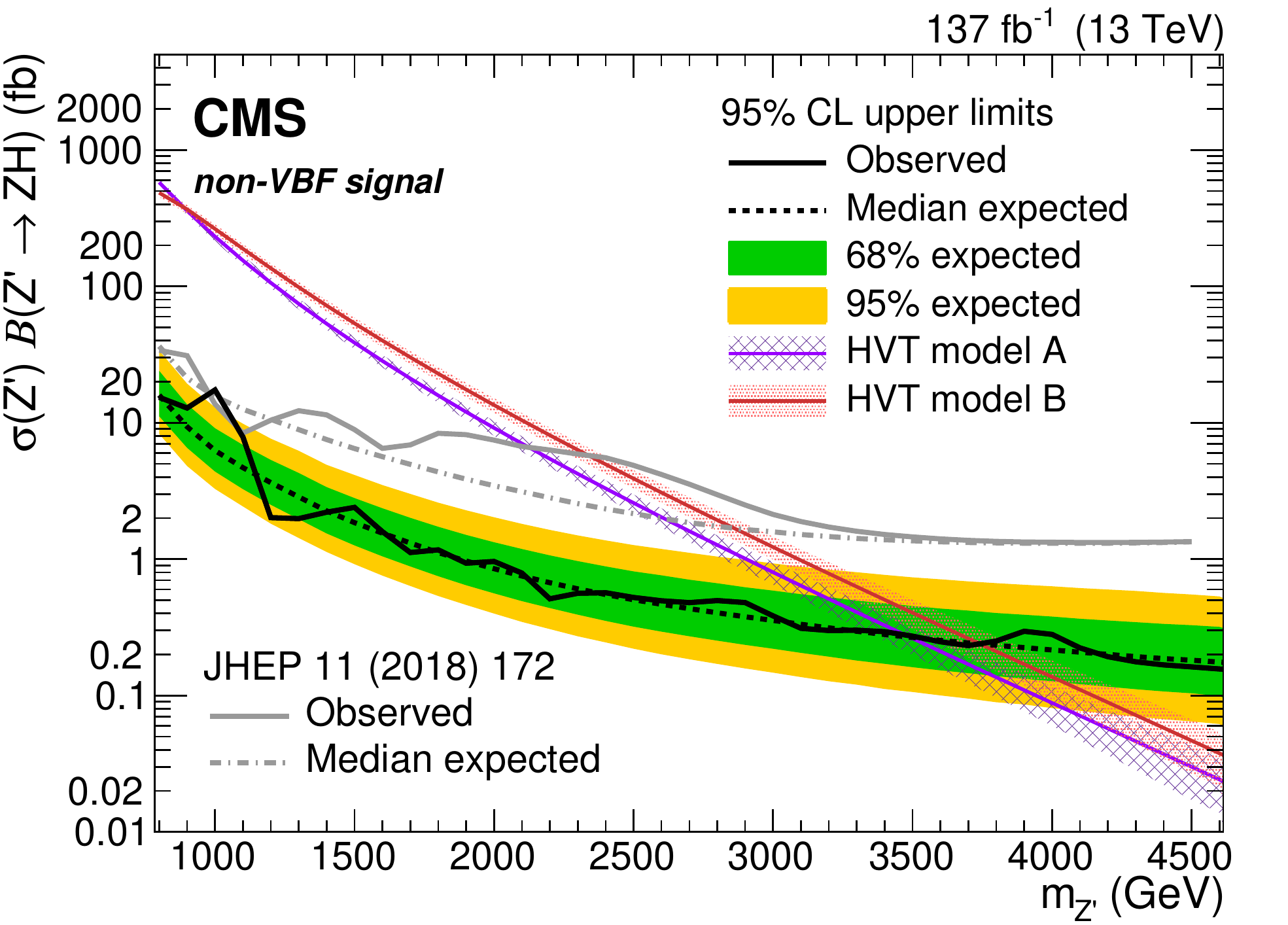}
  \includegraphics[width=.49\textwidth]{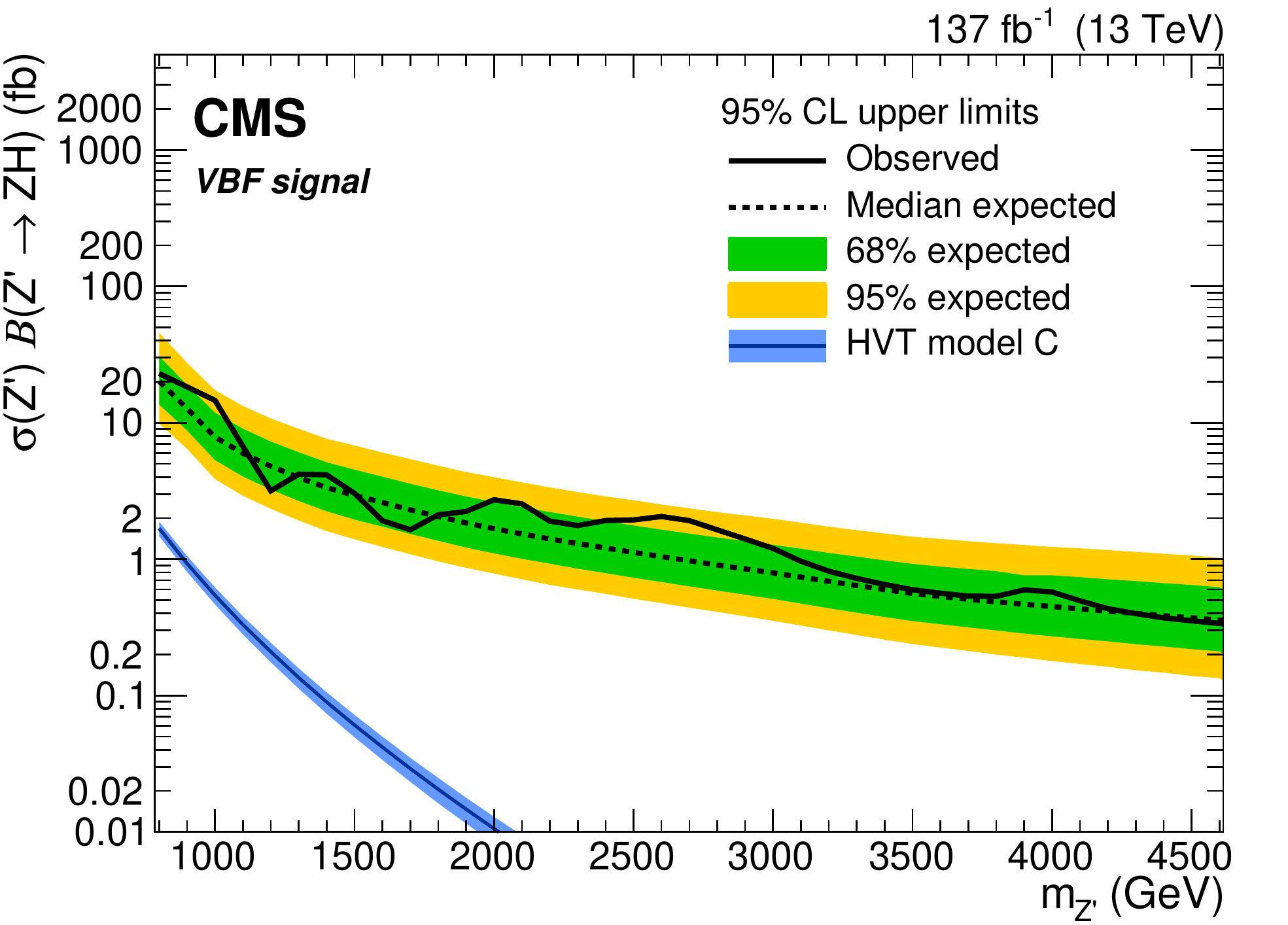}
  \caption{Observed and expected 95\% CL upper limit on $\sigma \B(\ZprtoZH)$ with all categories combined, for the non-VBF signal (\cmsLeft) and VBF signal (\cmsRight), including all statistical and systematic uncertainties. The inner green band and the outer yellow band indicate the regions containing 68 and 95\%, respectively, of the distribution of expected limits under the background-only hypothesis. The solid curves and their shaded areas correspond to the product of the cross section and the branching fractions predicted by the HVT models A and B (\cmsLeft) and HVT model C (\cmsRight), and their relative uncertainties. The CMS search for a heavy resonance using 2016 data and the same final state~\cite{1807.02826} is shown as a comparison.}
  \label{fig:Limit_XZHsl}
\end{figure*}

The exclusion limits on the product of resonance cross section and branching fraction $\B(\ZprtoZH)$ are reported as a function of the resonance mass in Fig.~\ref{fig:Limit_XZHsl} for all categories, separately for the non-VBF and the VBF signals. The $2\ell$ categories dominate the sensitivity for heavy resonance masses smaller than 1\TeV because of the smaller backgrounds combined with the better experimental resolution; at larger masses, the $0\ell$ categories are more sensitive thanks to the larger branching fraction of the \PZ boson to neutrinos. The exclusion limits are shown up to 4.6\TeV, which corresponds to the event with the highest \mX or \mtX observed either in the SB or SR.

The largest excess for the non-VBF signal, corresponding to a local significance of 3 standard deviations, is observed at $\mX=1\TeV$. A \PZpr boson with a mass smaller than 3.5\TeV is excluded at 95\% CL in HVT model A, and a \PZpr with mass smaller than 3.7\TeV is excluded in model B. The upper limit of the excluded mass range is increased by 0.85 (0.87)\TeV and 1.3 (1.4)\TeV) in HVT model A (model B) compared to searches using 2016 data and the same final state by the ATLAS and CMS Collaborations, respectively~\cite{1807.02826,1712.06518}. If the \PZpr couples only to the SM bosons and is produced exclusively through VBF as in HVT model C, the data set analyzed is not large enough to exclude any range of mass. Upper limits on the product of the cross section and branching fraction are set between 23 and 0.3\fb for a \PZpr mass between 0.8 and 4.6\TeV, respectively.

\begin{figure}[!htb]
  \centering
  \includegraphics[width=\cmsFigWidth]{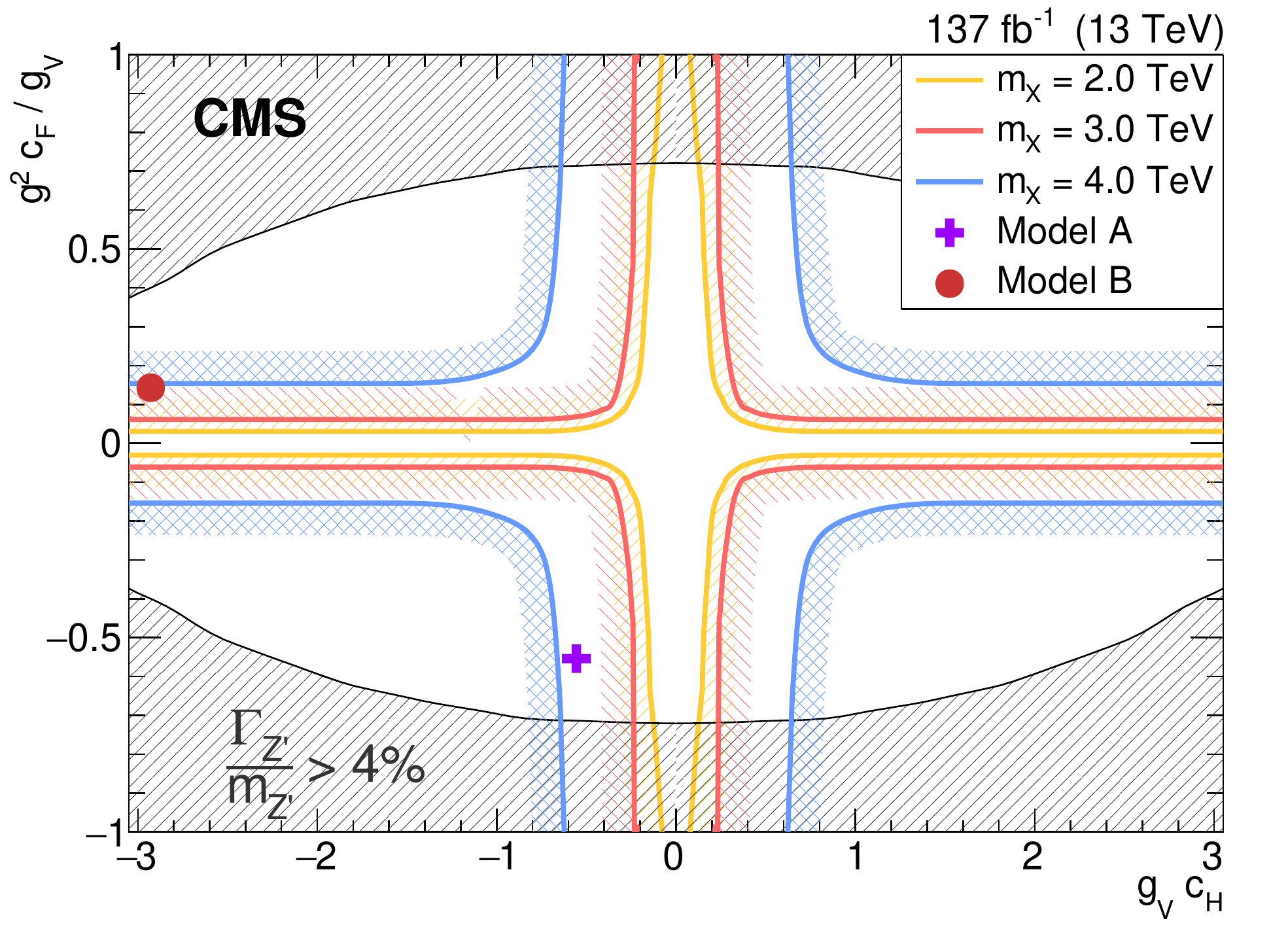}
  \caption{Observed exclusion limit in the space of the HVT model parameters [$\gV\cH$, $g^2\cF/\gV$], described in the text, for three different mass hypotheses of 2.0, 3.0, and 4.0\TeV for the non-VBF signal. The shaded bands indicate the side of each contour that is excluded. The benchmark scenarios corresponding to HVT models A and B are represented by a purple cross and a red point, respectively. The region of the parameter space where the natural resonance width ($\Gamma_{Z'}$) is larger than the typical experimental resolution of 4\%, for which the narrow-width approximation is not valid, is shaded in grey.}
  \label{fig:Parameterlimit_XZHsl}
\end{figure}

The exclusion limit of the non-VBF signal shown in Fig.~\ref{fig:Limit_XZHsl} (\cmsLeft) can be interpreted as a limit in the space of the HVT model parameters [$\gV\cH$, $g^2\cF/\gV$]. Combining all categories, the excluded region in such a parameter space for narrow resonances is shown in Fig.~\ref{fig:Parameterlimit_XZHsl}. The region of parameter space where the natural resonance width is larger than the typical experimental resolution of 4\%, for which the narrow width assumption is not valid, is shaded. 

\section{Summary}

A search for a heavy resonance with a mass between 0.8 and 5.0\TeV, decaying to a \PZ boson and a Higgs boson, has been described. The data samples were collected by the CMS experiment in the period 2016--2018 at $\sqrt{s}=13\TeV$ and correspond to an integrated luminosity of 137\fbinv. In the final states explored the \PZ boson decays leptonically, resulting in events with either zero or two electrons or muons. Higgs bosons with a large Lorentz boost are reconstructed via their decays to hadrons. For models with a narrow spin-1 resonance, a new heavy vector boson \PZpr with mass below 3.5 and 3.7\TeV is excluded at 95\% confidence level in models where the heavy vector boson couples predominantly to fermions and bosons, respectively. These are the most stringent limits placed on the Heavy Vector Triplet \PZpr model to date. If the heavy vector boson couples exclusively to standard model bosons, upper limits on the product of the cross section and branching fraction are set between 23 and 0.3\fb for a \PZpr mass between 0.8 and 4.6\TeV, respectively. This is the first limit set on a heavy vector boson coupling exclusively to standard model bosons in its production and decay.

\begin{acknowledgments}
  We congratulate our colleagues in the CERN accelerator departments for the excellent performance of the LHC and thank the technical and administrative staffs at CERN and at other CMS institutes for their contributions to the success of the CMS effort. In addition, we gratefully acknowledge the computing centres and personnel of the Worldwide LHC Computing Grid and other centres for delivering so effectively the computing infrastructure essential to our analyses. Finally, we acknowledge the enduring support for the construction and operation of the LHC, the CMS detector, and the supporting computing infrastructure provided by the following funding agencies: BMBWF and FWF (Austria); FNRS and FWO (Belgium); CNPq, CAPES, FAPERJ, FAPERGS, and FAPESP (Brazil); MES (Bulgaria); CERN; CAS, MoST, and NSFC (China); COLCIENCIAS (Colombia); MSES and CSF (Croatia); RIF (Cyprus); SENESCYT (Ecuador); MoER, ERC PUT and ERDF (Estonia); Academy of Finland, MEC, and HIP (Finland); CEA and CNRS/IN2P3 (France); BMBF, DFG, and HGF (Germany); GSRT (Greece); NKFIA (Hungary); DAE and DST (India); IPM (Iran); SFI (Ireland); INFN (Italy); MSIP and NRF (Republic of Korea); MES (Latvia); LAS (Lithuania); MOE and UM (Malaysia); BUAP, CINVESTAV, CONACYT, LNS, SEP, and UASLP-FAI (Mexico); MOS (Montenegro); MBIE (New Zealand); PAEC (Pakistan); MSHE and NSC (Poland); FCT (Portugal); JINR (Dubna); MON, RosAtom, RAS, RFBR, and NRC KI (Russia); MESTD (Serbia); SEIDI, CPAN, PCTI, and FEDER (Spain); MOSTR (Sri Lanka); Swiss Funding Agencies (Switzerland); MST (Taipei); ThEPCenter, IPST, STAR, and NSTDA (Thailand); TUBITAK and TAEK (Turkey); NASU (Ukraine); STFC (United Kingdom); DOE and NSF (USA).
  
  \hyphenation{Rachada-pisek} Individuals have received support from the Marie-Curie program and the European Research Council and Horizon 2020 Grant, contract Nos.\ 675440, 724704, 752730, and 765710 (European Union); the Leventis Foundation; the Alfred P.\ Sloan Foundation; the Alexander von Humboldt Foundation; the Belgian Federal Science Policy Office; the Fonds pour la Formation \`a la Recherche dans l'Industrie et dans l'Agriculture (FRIA-Belgium); the Agentschap voor Innovatie door Wetenschap en Technologie (IWT-Belgium); the F.R.S.-FNRS and FWO (Belgium) under the ``Excellence of Science -- EOS" -- be.h project n.\ 30820817; the Beijing Municipal Science \& Technology Commission, No. Z191100007219010; the Ministry of Education, Youth and Sports (MEYS) of the Czech Republic; the Deutsche Forschungsgemeinschaft (DFG), under Germany's Excellence Strategy -- EXC 2121 ``Quantum Universe" -- 390833306, and under project number 400140256 - GRK2497; the Lend\"ulet (``Momentum") Program and the J\'anos Bolyai Research Scholarship of the Hungarian Academy of Sciences, the New National Excellence Program \'UNKP, the NKFIA research grants 123842, 123959, 124845, 124850, 125105, 128713, 128786, and 129058 (Hungary); the Council of Science and Industrial Research, India; the HOMING PLUS program of the Foundation for Polish Science, cofinanced from European Union, Regional Development Fund, the Mobility Plus program of the Ministry of Science and Higher Education, the National Science Center (Poland), contracts Harmonia 2014/14/M/ST2/00428, Opus 2014/13/B/ST2/02543, 2014/15/B/ST2/03998, and 2015/19/B/ST2/02861, Sonata-bis 2012/07/E/ST2/01406; the National Priorities Research Program by Qatar National Research Fund; the Ministry of Science and Higher Education, project no. 0723-2020-0041 (Russia); the Programa Estatal de Fomento de la Investigaci{\'o}n Cient{\'i}fica y T{\'e}cnica de Excelencia Mar\'{\i}a de Maeztu, grant MDM-2015-0509 and the Programa Severo Ochoa del Principado de Asturias; the Thalis and Aristeia programs cofinanced by EU-ESF and the Greek NSRF; the Rachadapisek Sompot Fund for Postdoctoral Fellowship, Chulalongkorn University and the Chulalongkorn Academic into Its 2nd Century Project Advancement Project (Thailand); the Kavli Foundation; the Nvidia Corporation; the SuperMicro Corporation; the Welch Foundation, contract C-1845; and the Weston Havens Foundation (USA).
\end{acknowledgments}

\bibliography{auto_generated}

\cleardoublepage \appendix\section{The CMS Collaboration \label{app:collab}}\begin{sloppypar}\hyphenpenalty=5000\widowpenalty=500\clubpenalty=5000\vskip\cmsinstskip
\textbf{Yerevan Physics Institute, Yerevan, Armenia}\\*[0pt]
A.M.~Sirunyan$^{\textrm{\dag}}$, A.~Tumasyan
\vskip\cmsinstskip
\textbf{Institut f\"{u}r Hochenergiephysik, Wien, Austria}\\*[0pt]
W.~Adam, T.~Bergauer, M.~Dragicevic, J.~Er\"{o}, A.~Escalante~Del~Valle, R.~Fr\"{u}hwirth\cmsAuthorMark{1}, M.~Jeitler\cmsAuthorMark{1}, N.~Krammer, L.~Lechner, D.~Liko, I.~Mikulec, F.M.~Pitters, N.~Rad, J.~Schieck\cmsAuthorMark{1}, R.~Sch\"{o}fbeck, M.~Spanring, S.~Templ, W.~Waltenberger, C.-E.~Wulz\cmsAuthorMark{1}, M.~Zarucki
% ...
\vskip\cmsinstskip
\dag: Deceased\\
1:  Also at Vienna University of Technology, Vienna, Austria\\
% ...
\end{sloppypar}
\end{document}